\documentclass[fleqn,10pt]{olplainarticle}

\title{Test-beam performance results of the FASTPIX sub-nanosecond CMOS pixel sensor demonstrator}

\author[1,2,*]{Justus Braach}
\author[1]{Eric Buschmann}
\author[1]{Dominik Dannheim}
\author[1,3]{Katharina Dort}
\author[4]{Thanushan Kugathasan}
\author[4]{Magdalena Munker}
\author[1]{Walter Snoeys}
\author[1]{Peter \v{S}vihra}
\author[4]{Mateus Vicente}
\affil[1]{CERN, 1211 Geneva, Switzerland}
\affil[2]{Institut für Experimentalphysik, University of Hamburg, 20146 Hamburg, Germany}
\affil[3]{II. Physikalisches Institut, University of Giessen, 35390 Giessen, Germany}
\affil[4]{DPNC, University of Geneva, 1205 Geneva, Switzerland}
\affil[*]{Correspondence: justus.braach@cern.ch}

\keywords{monolithic pixel sensor, particle detection, fast timing}

\begin{abstract}
Within the ATTRACT FASTPIX project, a monolithic pixel sensor demonstrator chip has been developed in a modified \SI{180}{\nm} CMOS imaging process technology, targeting sub-nanosecond timing precision for single ionising particles. 
It features a small collection electrode design on a \SI{25}{\um}-thick epitaxial layer and contains 32 mini matrices of 68 hexagonal pixels each, with pixel pitches ranging from \num{8.66} to \SI{20}{\um}. 
Four pixels are transmitting an analog output signal and 64 are transmitting binary hit information. 
Various design variations are explored, aiming at accelerating the charge collection and making the timing of the charge collection more uniform over the pixel area. 
Signal treatment of the analog waveforms, as well as reconstruction of digital position, time and charge information, is carried out off-chip. 
This contribution introduces the design of the sensor and readout system and presents performance results for various pixel designs achieved in recent test-beam measurements with external tracking and timing reference detectors. 
A time resolution below \SI{150}{\ps} is obtained at full efficiency for all pixel pitches.
\end{abstract}

\begin{document}

\flushbottom
\maketitle
\thispagestyle{empty}

\section{Introduction}\label{sec:intro}

Vertex and tracking detectors for future high-energy physics experiments face stringent requirements in view of their spatial and temporal measurement performance as well as the projected experimental conditions. 
Within the ATTRACT FASTPIX project and under the umbrella of CERN’s strategic research and development program for future experiments (EP R\&D) \cite{Aleksa:2649646}, a monolithic pixel sensor demonstrator chip has been developed in a modified \SI{180}{\nm} CMOS imaging process technology \cite{ATTRACTFASTPIX}. It targets sub-nanosecond time tagging of hits from single minimum-ionizing particles, using small pixels of \SI{\leq 20}{\um} pitch.\\
For FASTPIX small line width CMOS technologies allow the design of hexagonal pixels with few-femtofarad collection electrodes and large signal-to-noise ratios, in favor of detection efficiency and precise timing. 
Low-field regions, non-uniform electric field configurations and space requirements for complex in-pixel circuits introduce timing variation depending on the in-pixel particle incidence location. 
In order to face a tradeoff between pixel size and non-uniform timing, design variations are introduced to accelerate the propagation of the signal charge and decrease the in-pixel timing variation. 
The development of CMOS sensors with small collection electrode design using Technology Computer Aided Design (TCAD) simulations was presented in \cite{Munker2019}. 
The sensor-design optimization and the circuit design supporting sensor timing performance in the sub-nanosecond range is presented in \cite{KUGATHASAN2020164461}.\\
After first results from measurements with two matrices of a 
sample chip presented in \cite{instruments6010013}, this publication accesses a larger set of measurement data for a comparison of the various process modifications, sensor layouts and pixel pitches. 
The aim is to investigate the potential of targeted sensor process modifications and design optimizations for the detector performance.

\section{The FASTPIX chip}\label{sec:FASTPIX}
\subsection{Sensor design}\label{sec:sensor_design}
\begin{figure}[ht]
    \centering
    \includegraphics[width=\linewidth]{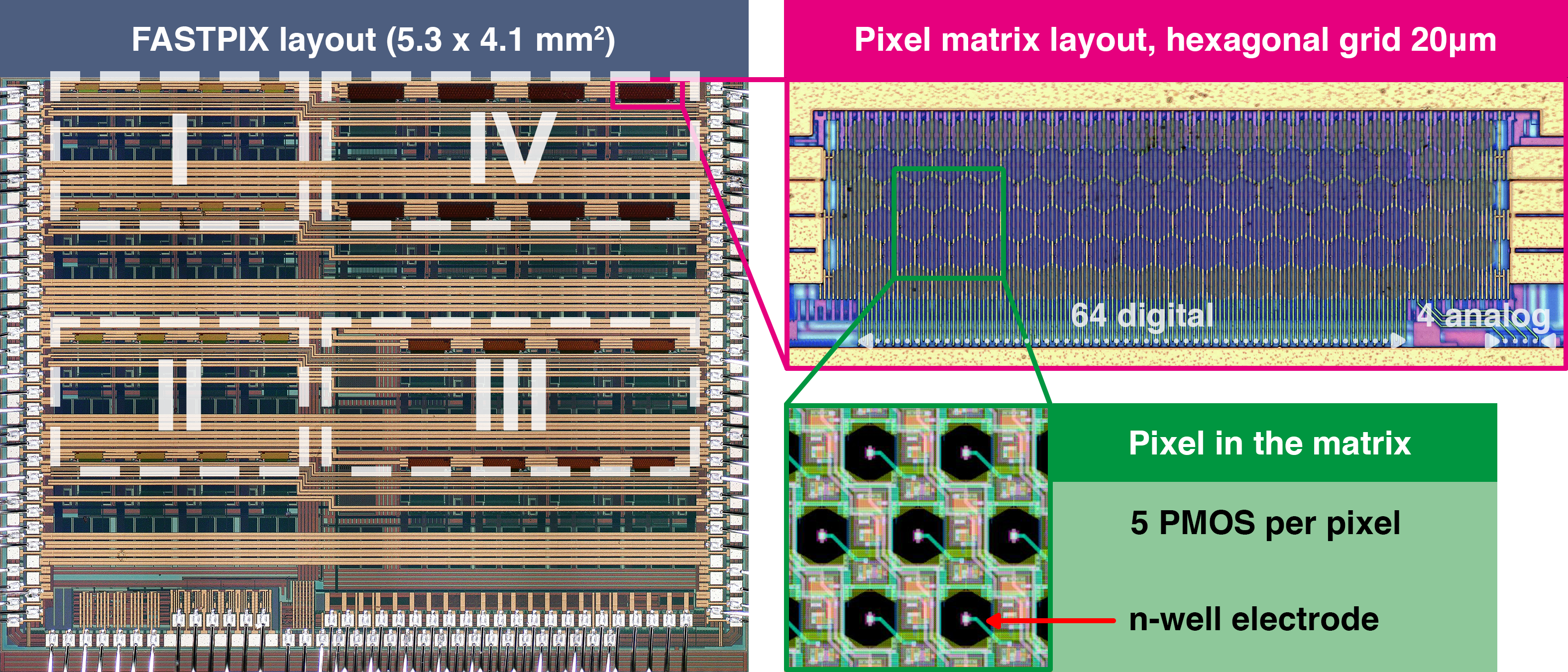}
    \caption{Photograph of the FASTPIX chip (left), divided in four quadrants of matrices with pixel pitches I: \SI{8.66}{\um}, II: \SI{10}{\um}, III: \SI{15}{\um} and IV: \SI{20}{\um}. An exemplary picture and zoomed in details of a \SI{20}{\um} pitch matrix (right). 
    Modified from \cite{instruments6010013}.} 
    \label{fig:chip_overview}
\end{figure}
The layout of the FASTPIX chip shown in \cref{fig:chip_overview} is partitioned into four groups of eight matrices each, with pixel pitch I: \SI{8.66}{\um}, II: \SI{10}{\um}, III: \SI{15}{\um} and IV: \SI{20}{\um}.
\begin{figure}[!h]
    \centering
    \includegraphics[width=\linewidth]{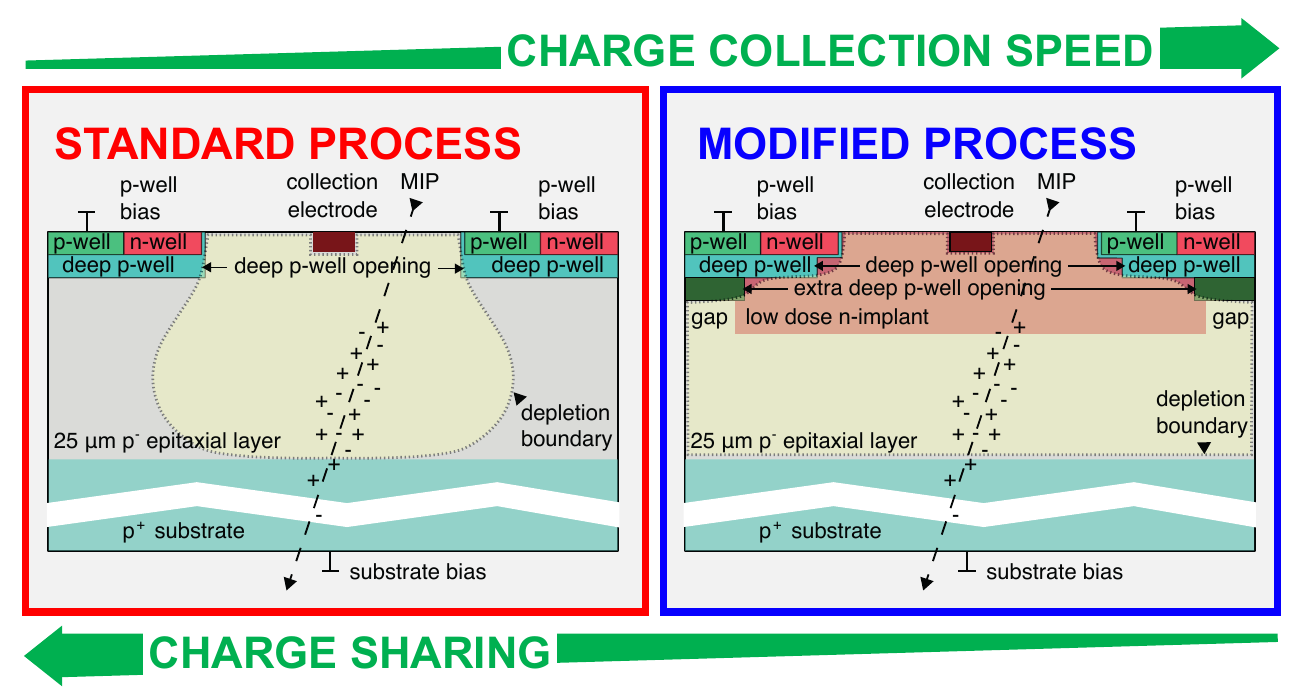}
    \caption{Wafer production process variants for FASTPIX represented by schematic cross-sections of the pixel unit cells, showing a cut perpendicular to the sensor surface. The standard \SI{180}{\nm} CMOS imaging process (left) and the modified process variant (right) with added low-dose n-type implant and optimizations such as a gap in the n-implant, retracted deep p-well and additional extra-deep p-well implant.} 
    \label{fig:process_variants}
\end{figure}
The pixel pitch is the radial distance between two adjacent collection electrodes or equivalent to the diameter of the incircle of a hexagonal pixel.
The sensor has been produced in multiple versions that are all based on two variants of the manufacturing process illustrated in \cref{fig:process_variants}. 
The standard process sets the baseline with no additional measures to accelerate the collection of signal charge and with incomplete lateral depletion of the sensor volume.
A timing spread is introduced by a difference in charge collection time for a hit in the corner compared to a hit in the center of the pixel. 
Minimizing the maximum distance between charge generation and collection is a first step in the optimization for timing uniformity and is realized by the hexagonal arrangement of collection electrodes and the $\mathcal{O}(\SI{10}{\um})$ pixel pitch. 
The geometry also has an effect on charge sharing at the pixel corners with only three neighboring pixels instead of four, as it is the case for rectangular geometries.
With the modified process, a uniform low-dose n-type implant is introduced allowing full lateral depletion of the \SI{25}{\um} epitaxial layer.\\ 
In production \num{12} process splits were implemented from combinations of different doping concentration and two different versions of the two lithographic photomasks that control the feature sizes of n-type and p-type implantation into the epitaxial layer.
Among those, optimizations geared towards improved timing performance feature an additional extra-deep p-well implant as well as a gap in the n-implant near the pixel edge.
\begin{figure}[b!]
    \centering
    \includegraphics[width=\linewidth]{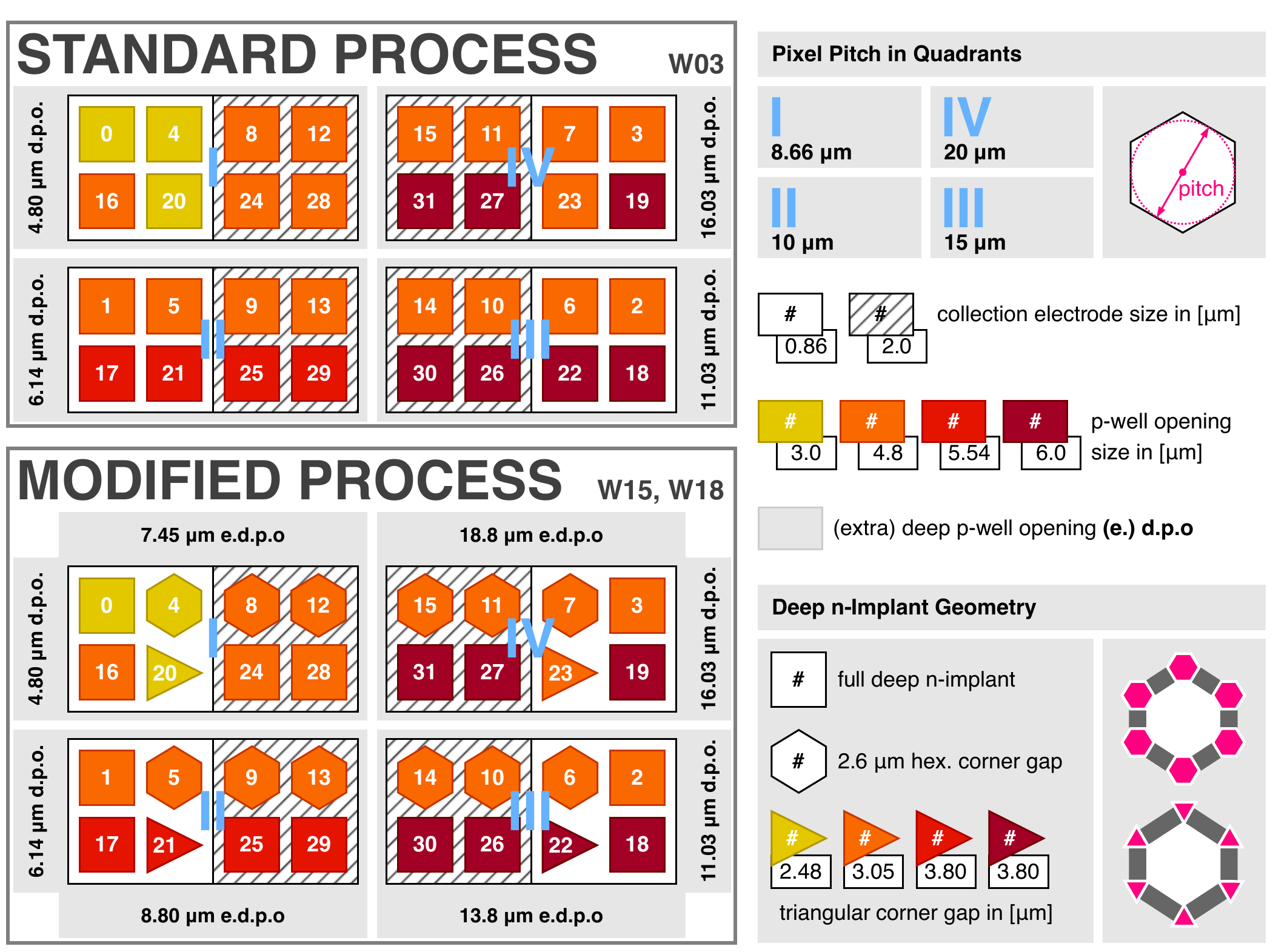}
    \caption{Overview of 32 matrices implemented on a single FASTPIX chip. The top-left scheme illustrates the parameter space of wafers produced using the standard process, below a similar representation of \numproduct{4 x 8} matrices on a chip produced in the modified process. The four quadrants are indicated by light blue roman numeral and are attributed their pixel pitch in the table on the top right. Within each quadrant two different collection electrode sizes are represented by white or hatched background. The matrices themselves are illustrated by geometrical icons colored in yellow, orange, red, or dark-red with the respective matrix number in bold white. The colorway indicates the p-well opening size of the pixel cell. The deep p-well opening and extra-deep p-well opening size is written along the outside edges of each quadrant. In case of the modified process different deep n-implant (gap) geometries are represented by the shape of the icon of a given matrix.} 
    \label{fig:parameter_space}
\end{figure}
Both measures introduce a gradient in the doping profile and with that increase the strength of the lateral electric field in the corner regions of the pixel.
A more pronounced pull away from the pixel border towards the collection electrode accelerates charge collection by drift, reduces charge sharing, maximizes the seed pixel signal and narrows the spread in timing performance.\\
\cref{fig:parameter_space} gives an overview of the large parameter space of layout variations implemented across the 32 matrices on a single FASTPIX chip. The following sections will discuss FASTPIX samples of three wafers: W03 (standard process), W15 (modified process), W18 (modified process with higher-dose deep n-implant).

%
\subsection{Frontend and readout architecture}\label{sec:frontend_readout}
The 68 active pixels of a given matrix are grouped into 64 digital and 4 analog channels that each have in-pixel circuitry for reset and leakage current compensation and one stage of amplification. 
Due to space constraints in the \SI{180}{\nm} CMOS imaging process, the second amplification stage is moved to the matrix periphery.
Analog channels are routed through buffers while the signals for digital channels are processed by a discriminator block with fast-OR logic and two delay lines offering a time-based encoding of position, time and energy of a hit. 
The on-chip circuitry is capable to resolve small differences in charge collection time to support a detailed characterization of the timing performance of the sensor. 
Fast signals need to be buffered to reach the matrix periphery where the discriminators are implemented, which imposes a penalty on power consumption.
With \SI{\approx 20}{\mW} per pixel the circuit reaches a simulated time jitter of \SI{\approx 20}{\ps} \cite{KUGATHASAN2020164461}.
The three low voltage differential output signals of the chip are read out asynchronously and get processed further off-chip. 

\section{Measurement Setup}\label{sec:setup}
The chip is mounted on a carrier printed circuit board (PCB) which receives power, bias and control from the Caribou readout system \cite{Vanat:2020nS}.
An oscilloscope with $\mathcal{O}(\si{\giga\hertz})$ analog bandwidth and a sampling rate of $\mathcal{O}(\SI{10}{\giga\siemens\per\second})$ is also controlled by Caribou and used to record the digital channel signals.
\\
For the test beam measurements presented in the following sections the setup is installed in the CLICdp Timepix3 beam telescope at the SPS North Area at CERN \cite{DallOcco:2651308, AlipourTehrani:2270788, AbuslemeHoffman:429703}.
FASTPIX samples are positioned on a x-/y-/rotation-stage in the center of the telescope between three upstream and three downstream Timepix3 reference planes.
The FASTPIX device under test (DUT) is operated at \SI{-6}{\volt} bias at the substrate and p-well electrodes.
The Micro-Channel-Plate Photomultiplier Tube (MCP-PMT) HPK R3809U-50 \cite{R3809U} provides reference time measurements \cite{BORTFELDT2020163592} and is connected to a fourth channel of the oscilloscope. 
To minimize the amount of material in the telescope acceptance, the MCP-PMT is positioned after the downstream set of telescope planes and is biased at \SI{2.6}{\kilo\volt} resulting in a reference track time resolution of \SI{< 10}{\ps}.\\
Several FASTPIX samples have been tested in dedicated high-intensity \SI[per-mode=symbol]{120}{\giga\electronvolt\per\light} pion beam periods in 2022. 
The small active area of the matrices typically yield between \num{100} and \num{500} hits per \SI{4.8}{\s} spill, resulting in average rates of \qtyrange[range-phrase = \:--\:]{5}{25}{\Hz}.
In preparation for test beam measurement campaigns, FASTPIX samples have been commissioned, tested and calibrated following the methodology presented in \cite{instruments6010013}.
To allow a good performance comparison between matrices, the threshold for each matrix has been calibrated by externally injecting electrical test pulse signals into two digital pixels of a given matrix. 

\section{Reconstruction and Analysis}\label{sec:analysis}
Offline reconstruction and analysis of the test-beam data is performed using the software framework Corryvreckan \cite{Dannheim2021}, with extensions for the time-based decoding of position and pulse height as well as the proper treatment of the hexagonal pixel geometry.

\subsection{Data decoding}\label{sec:decoding}
For FASTPIX the recorded raw waveforms require offline decoding \cite{instruments6010013}. 
The time-of-arrival (ToA) is calculated from the fast-OR output of the chip providing the first edge of the fastest pixel signal in a given matrix. 
The time-over-threshold (ToT) of a pixel signal follows as the distance between a pair of rectangular pulses. 
The discriminator outputs a pulse whenever the pixel signal rises above or falls below threshold.
The pixel position is obtained on the basis of a calibration of the FASTPIX delay lines.\\ 
The temporal positions of the \num{64} pixels within a delay line are determined from a calibration of a given matrix using single-pixel hits. 
For decoding, the coordinates of the spatial pixel center are attributed to the designed order of pixels in the delay line. 
The spatial position of a recorded pixel hit is then found by mapping the measured value to the calibrated pixel delays on the delay line. 
Overlap between signals from pixels simultaneously crossing the threshold poses a challenge for this pixel position en-/decoding scheme.
In case of signal overlap the correct position can still be identified, as long as the relative majority of signal combinations from both delay lines indicate the correct pixel position value. 
Hits that can not be completely decoded are discarded in the data decoding stage. 

\subsection{Event building}\label{sec:event_building}
The reference telescope planes provide spatial as well as temporal information for pixel hits and are read out continuously and independently from the DUT.
The oscilloscope triggers on a fast logic OR combination of all pixels in a FASTPIX matrix, providing the trigger information which is used to synchronize DUT and telescope planes. 
The analysis of observables in the spatial and temporal domain is based on telescope events that are linked to a trigger on FASTPIX. 
For a study of detector efficiency the telescope data stream is divided into \SI{10}{\us} intervals of all telescope events, independent of FASTPIX triggers. 
The time intervals are sized to avoid large hit multiplicities leading to processing intensive amount of combinations in subsequent reconstruction steps.

\subsection{Clustering}\label{sec:clustering}
Direct neighboring pixels collecting charge within a \SI{200}{\ns} time window are clustered 
and the pixel with the earliest timestamp is used to set the cluster timestamp.
Clusters get attributed a center position that is used to reconstruct the hit position.
In case of a single-pixel cluster the hit position is obtained by using the pixel center.
For cluster sizes $> 1$ a charge-weighted interpolation of the cluster geometry is applied, corrected for non-linear charge sharing using an $\eta$-correction algorithm \cite{BELAU1983253,AKIBA201231}.
For FASTPIX this correction is conducted in radial coordinates (r,$\phi$) to account for the different symmetry axes of the hexagonal pixel grid \cite{Wennloef2023}. 


\subsection{Tracking}\label{sec:tracking}
Straight line tracks are interpolated from combinations of clusters in telescope planes, as implemented in Corryvreckan \cite{Dannheim2021}. 
A track is created as soon as clusters from a minimum \num{6} telescope planes have been associated to a particle trajectory, given they are within a set of spatial (\SI{400}{\um} in projected x-/y-distance) and temporal cuts (\SI{20}{\ns}).
Cluster information from the device under test (DUT) is still left out at this point in the reconstruction in order to avoid a bias on the analysis.


\subsection{DUT analysis}\label{sec:FASTPIX_analysis}
Tracks are matched to DUT clusters if they occur within a \SI{20}{\ns} time window and if the distance between cluster center and track impact position is smaller than the pixel pitch of the investigated matrix. 
The matching is performed based on the distance between the cluster center and interpolated track intersect.
The analysis is set to exclude telescope tracks with $\nicefrac{\chi^2}{\text{ndf}}> 5$, at the event edges (\SI{250}{\ns}), during oscilloscope dead time (\SI{5}{\us}), while copying data from the buffer to the measurement server or close to the oscilloscope frame edge (\SI{> 250}{\ns}).
To avoid edge effects on the border of the pixel matrix, especially for cluster sizes $>$ \num{4} in the 4 x 16 pixel matrices, clusters with track intersect outside of the central 2 x 14 pixels are discarded.

\subsection{Timewalk correction}\label{sec:timewalk_corr}
\begin{figure}[!ht]
    \centering
    \begin{subcaptionblock}[t]{0.495\textwidth}
        \centering
        \includegraphics[width=.91\textwidth]{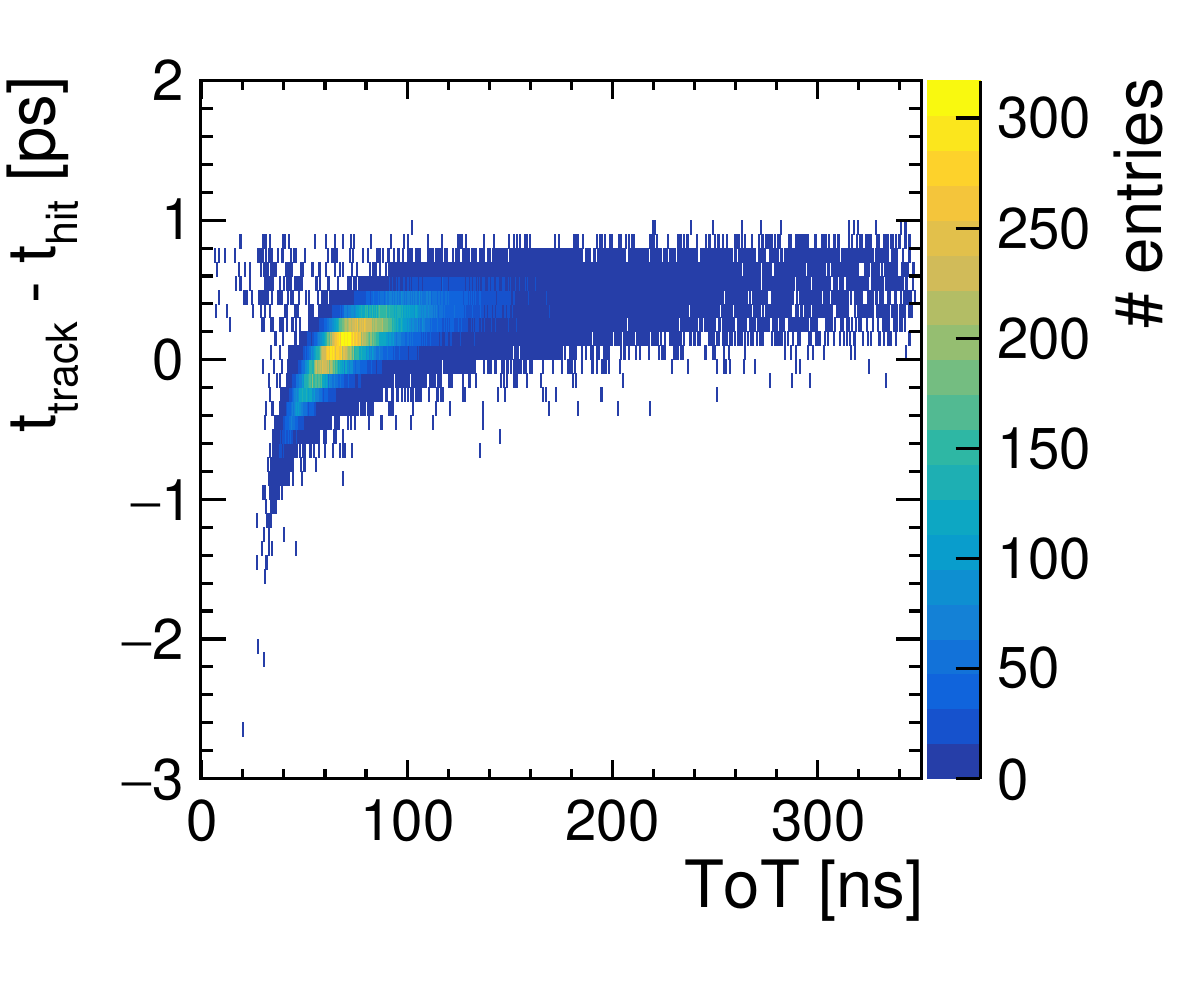}
        \caption{Without correction.}
        \label{fig:timewalk_2D_nocorr}
    \end{subcaptionblock}
    \hfill
    \begin{subcaptionblock}[t]{0.495\textwidth}
        \centering
        \includegraphics[width=.91\textwidth]{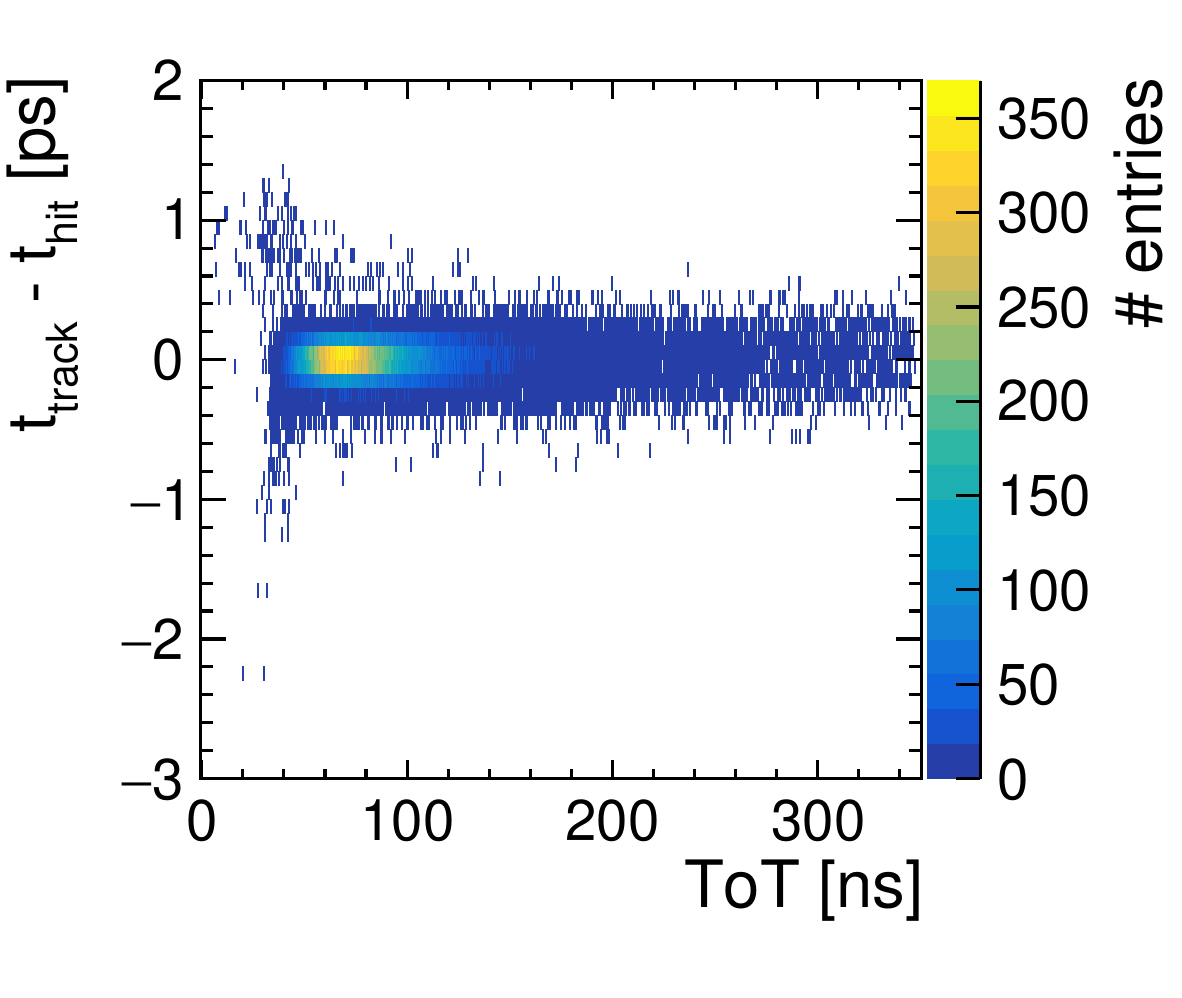}
        \caption{With timewalk correction.}
        \label{fig:timewalk_2D_corr}
    \end{subcaptionblock}
    \caption{Time residuals as function of seed signal for all pixels in the inner \num{2} x \num{14} pixel region of a \SI{20}{\um} pitch matrix.}
    \label{fig:timewalk_2D}
\end{figure}

For studies of time measurement precision, the fastest-signal pixel in a cluster is used to calculate the time residual with respect to the time reference.
\cref{fig:timewalk_2D_nocorr} shows the residuals over seed-pixel ToT for all pixels in the inner \num{2} x \num{14} pixel region of a FASTPIX matrix. 
\begin{figure}[!ht]
    \centering
    \begin{subcaptionblock}[t]{0.495\textwidth}
        \centering
        \includegraphics[width=.91\textwidth]{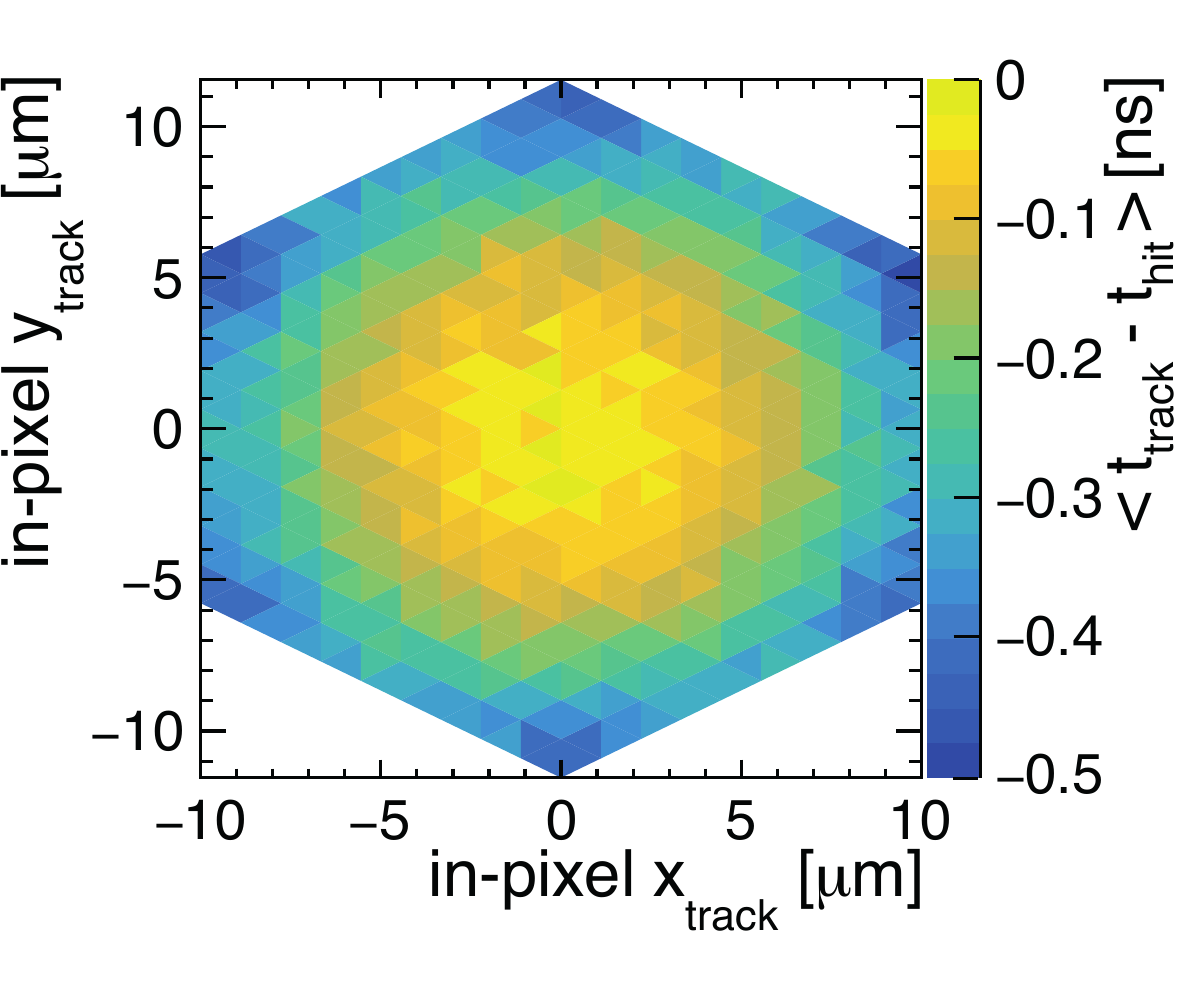}
        \caption{Mean time residual in relation to in pixel track position.}
    \end{subcaptionblock}
    \hfill
    \begin{subcaptionblock}[t]{0.495\textwidth}
        \centering
        \includegraphics[width=.91\textwidth]{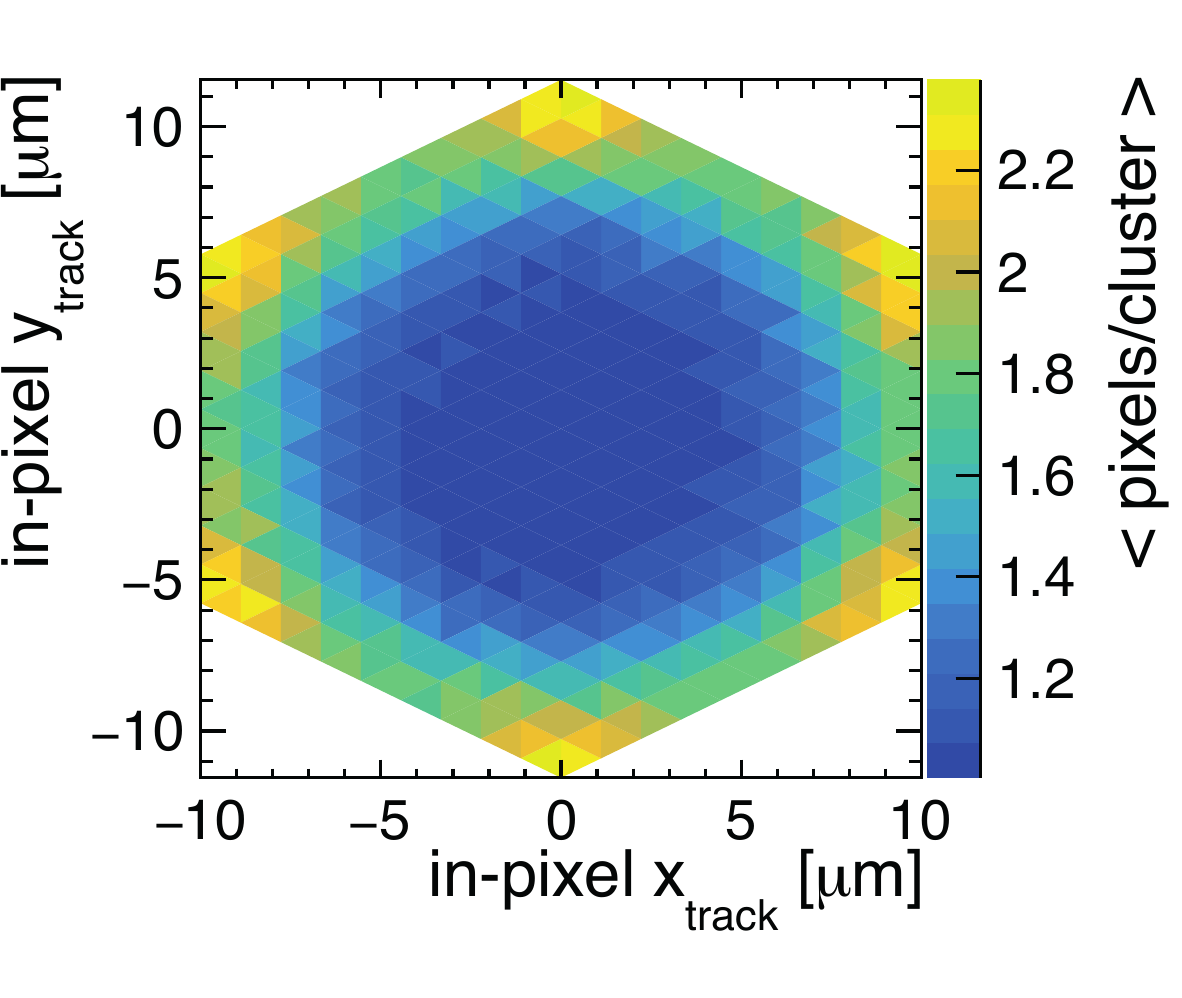}
        \caption{Mean number of pixels per cluster in relation to in pixel track position.}
    \end{subcaptionblock}
    \caption{Time residual and cluster-size maps for in-pixel track incidence positions in x-/y-plane of a \SI{20}{\um} pixel pitch matrix. Hits from all pixels are projected into one hexagonal pixel area.}
    \label{fig:timewalk_in-pixel}
\end{figure}
\\
The equivalent in-pixel plot and a related in-pixel cluster-size distribution is shown in \cref{fig:timewalk_in-pixel} and presents a dependency of the timing performance and correlated cluster size on the in-pixel particle incidence location.
This suggests a time-walk correction approach per cluster size applied on subsets of data.
For each cluster size \num{<= 5} an exclusive data set is assembled and used to evaluate the time residual in relation to ToT on a pixel-by-pixel level. 
Clusters \num{> 5} are grouped into a combined set to retain sufficient statistics for further analysis.
The correction is obtained by calculating the mean of each ToT bin of one half of a data set and subtracting said mean from the corresponding entries in the other half
, and vice versa.\\
\cref{fig:timewalk_2D_corr} shows the residuals over seed-pixel ToT for all pixels in the inner region of a FASTPIX matrix after time-walk correction. 
The cluster-size-specific subsets have been rejoined after individual correction and are presented together in a single plot.\\
\\
\newpage

\section{Results}\label{sec:results}
\subsection{Cluster size and efficiency}\label{sec:clstrsz_eff}
\begin{figure}[ht]
    \centering
    \includegraphics[width=.5\linewidth]{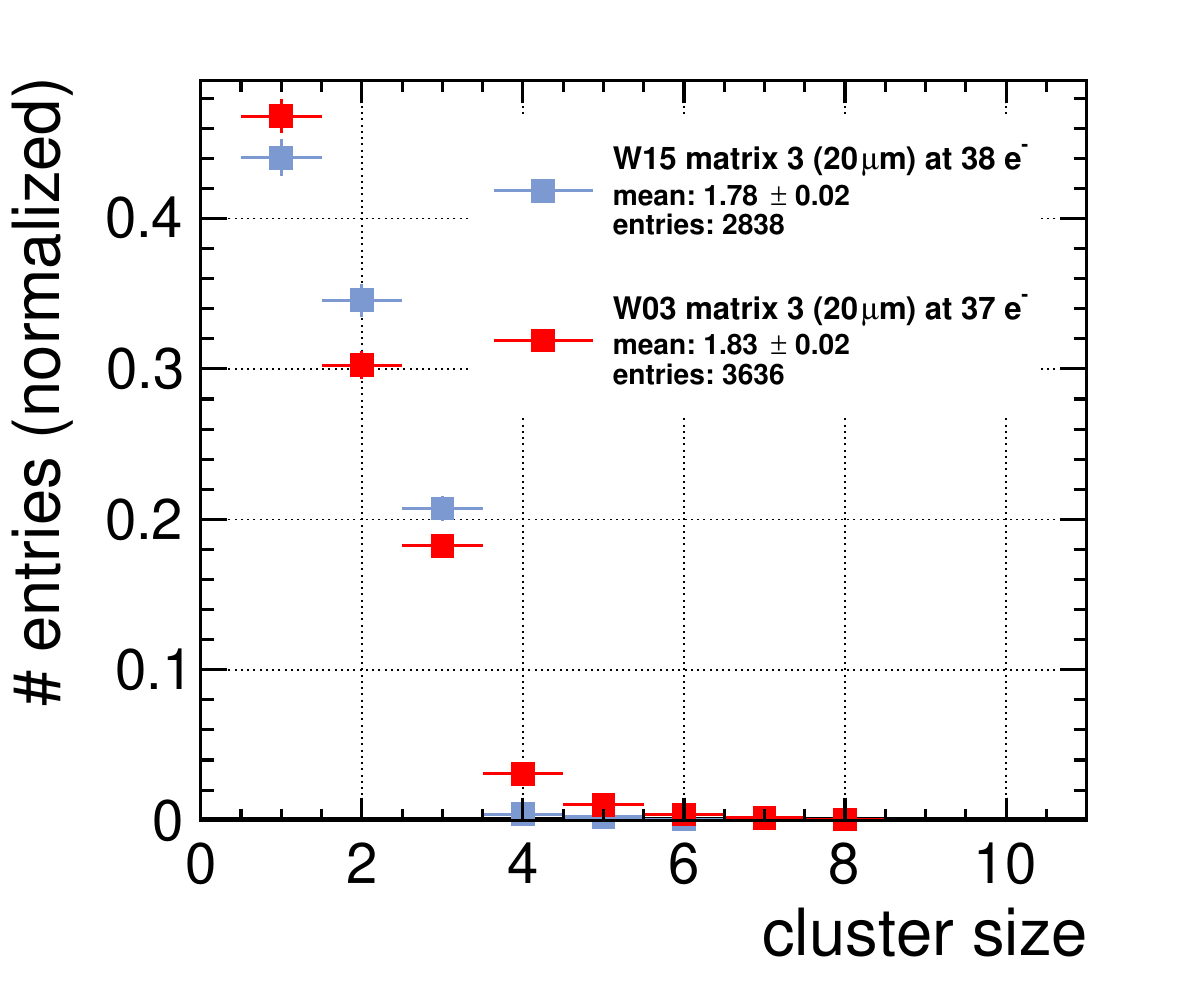}
    \caption{Cluster-size distributions of a baseline \SI{20}{\um} pixel pitch matrix on W03 and W15.} 
    \label{fig:clstrsz_dist_20}
\end{figure}
\begin{figure}[h!]
    \centering
    \begin{subcaptionblock}[t]{0.495\textwidth}
        \centering
        \includegraphics[width=\textwidth]{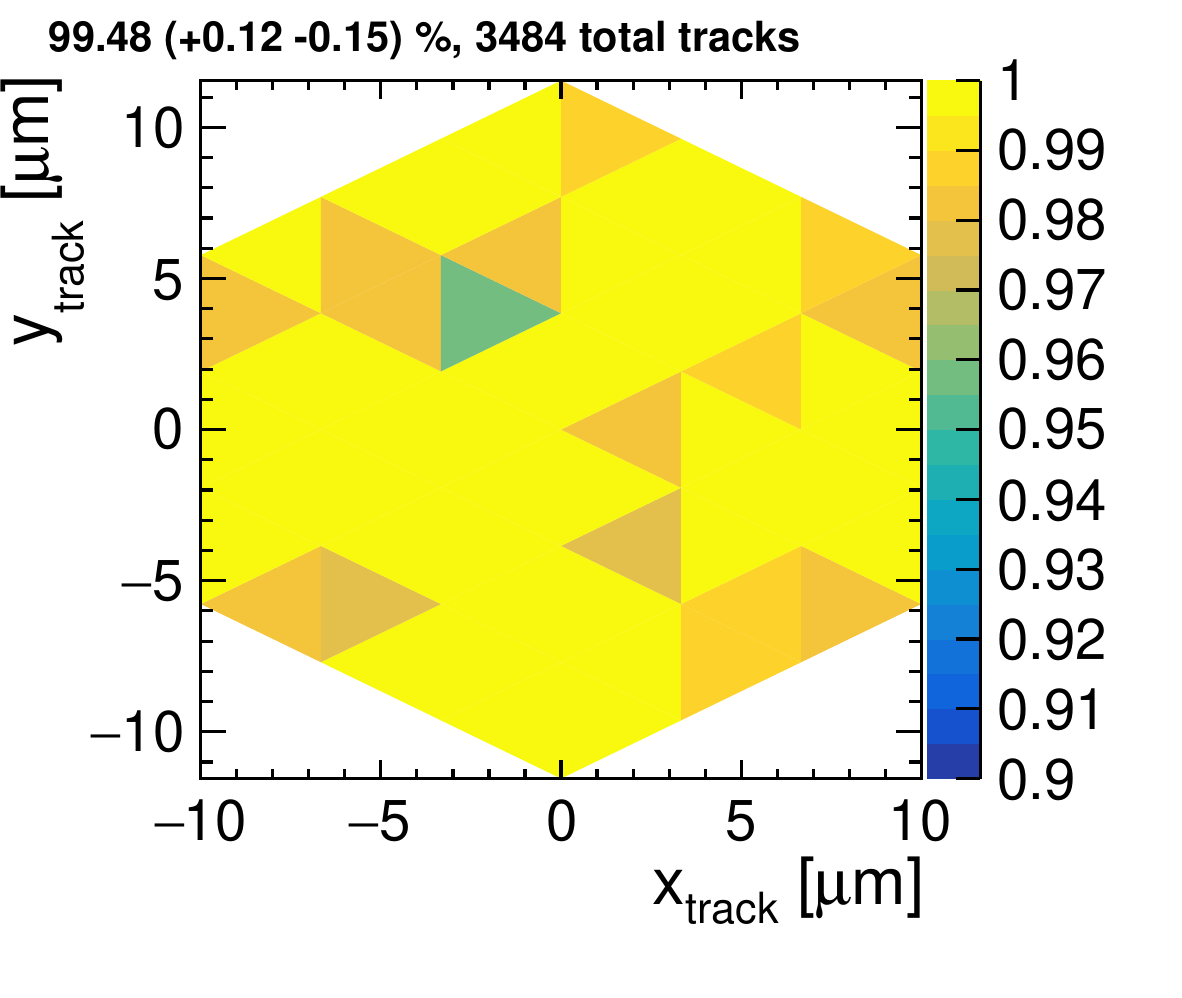}
        \caption{W03, matrix 0, \SI{20}{\um} pixel pitch, threshold \SI{38}{\electron}.}
    \end{subcaptionblock}
    \hfill
    \begin{subcaptionblock}[t]{0.495\textwidth}
        \centering
        \includegraphics[width=\textwidth]{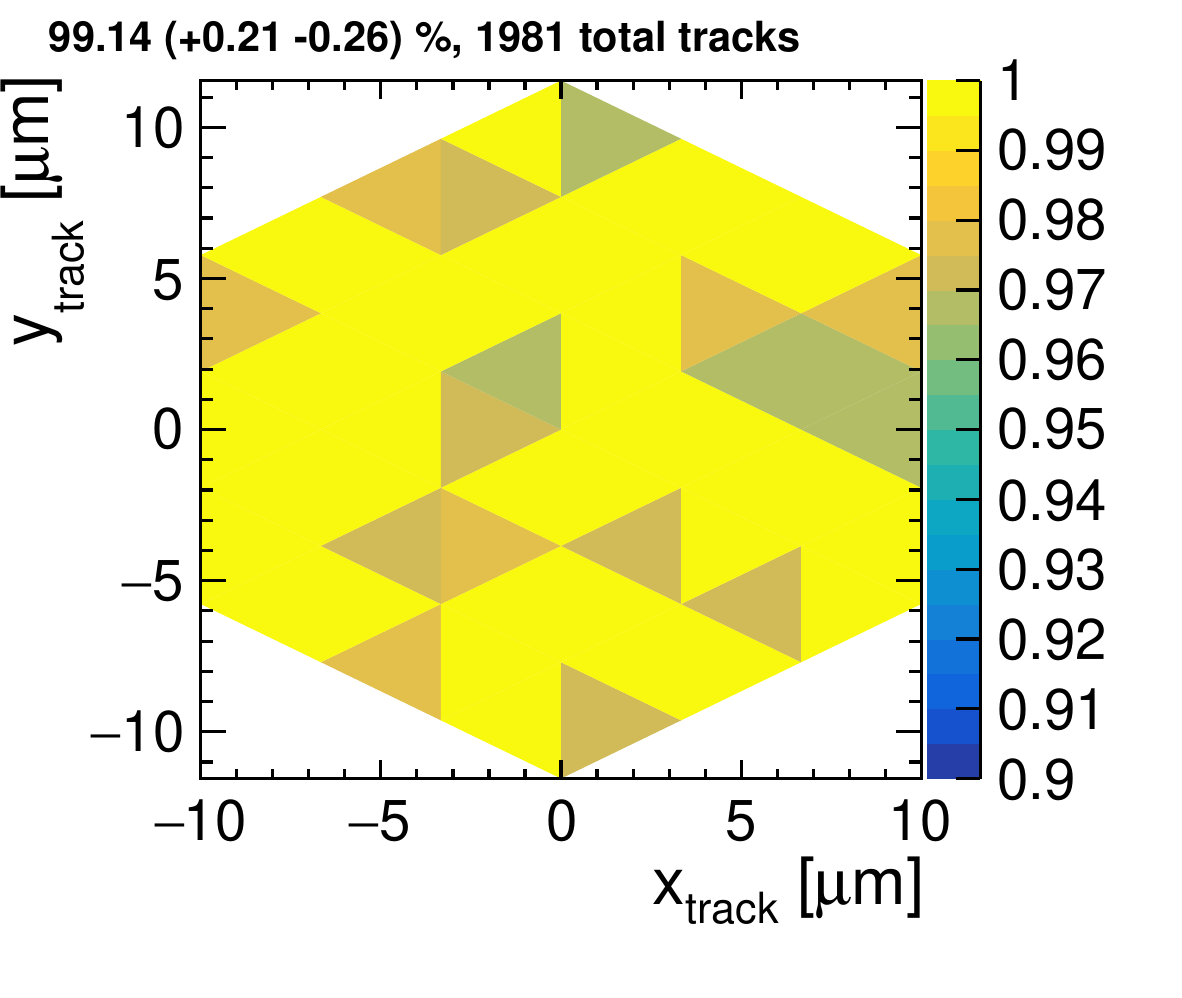}
        \caption{W15, matrix 0, \SI{20}{\um} pixel pitch, threshold \SI{38}{\electron}.}
    \end{subcaptionblock}
    \caption{In-pixel efficiency maps from a standard process (left) and a modified process (right) sample. Hits from all pixels are projected into one hexagonal pixel area. The color scale shows the efficiency for each triangular track position bin subdividing the hexagonal area on the x-/y-plane.}
    \label{fig:inpix_efficiency_20}
\end{figure}
\begin{figure}[h!]
    \centering
    \begin{subcaptionblock}[t]{0.495\textwidth}
        \centering
        \includegraphics[width=\textwidth]{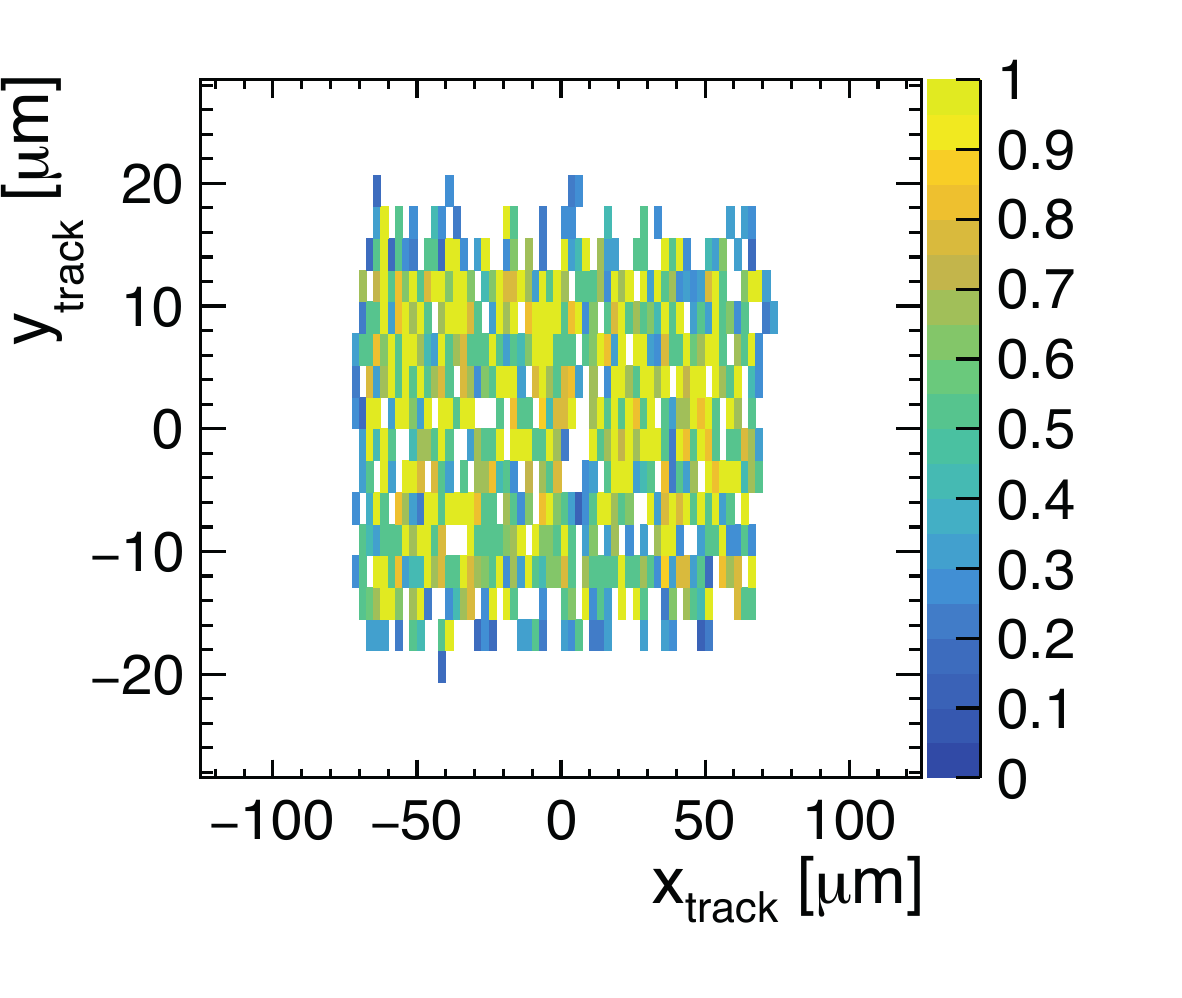}
        \caption{W03, matrix 3, \SI{8.66}{\um} pixel pitch, threshold \SI{66}{\electron}.}
    \end{subcaptionblock}
    \hfill
    \begin{subcaptionblock}[t]{0.495\textwidth}
        \centering
        \includegraphics[width=\textwidth]{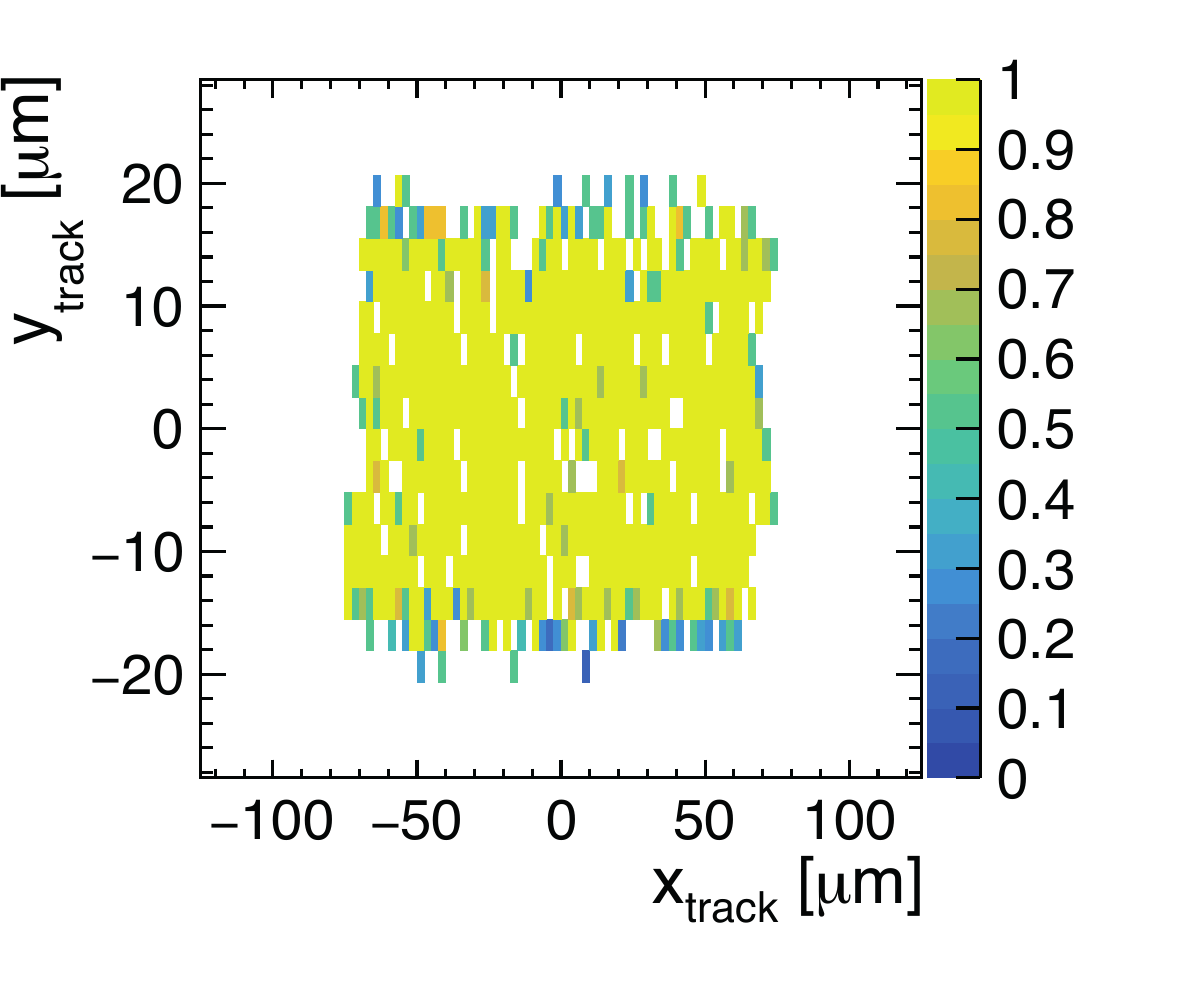}
        \caption{W15, matrix 3, \SI{8.66}{\um} pixel pitch, threshold \SI{51}{\electron}.}
    \end{subcaptionblock}
    \caption{Matrix efficiency maps from a standard process (left) and a modified process (right) sample. The color scale shows the efficiency for each track position bin on the x-/y-plane.}
    \label{fig:matrix_efficiency_8}
\end{figure}
\begin{figure}[t]
    \centering
    \begin{subcaptionblock}[t]{0.495\textwidth}
        \centering
        \includegraphics[width=\textwidth]{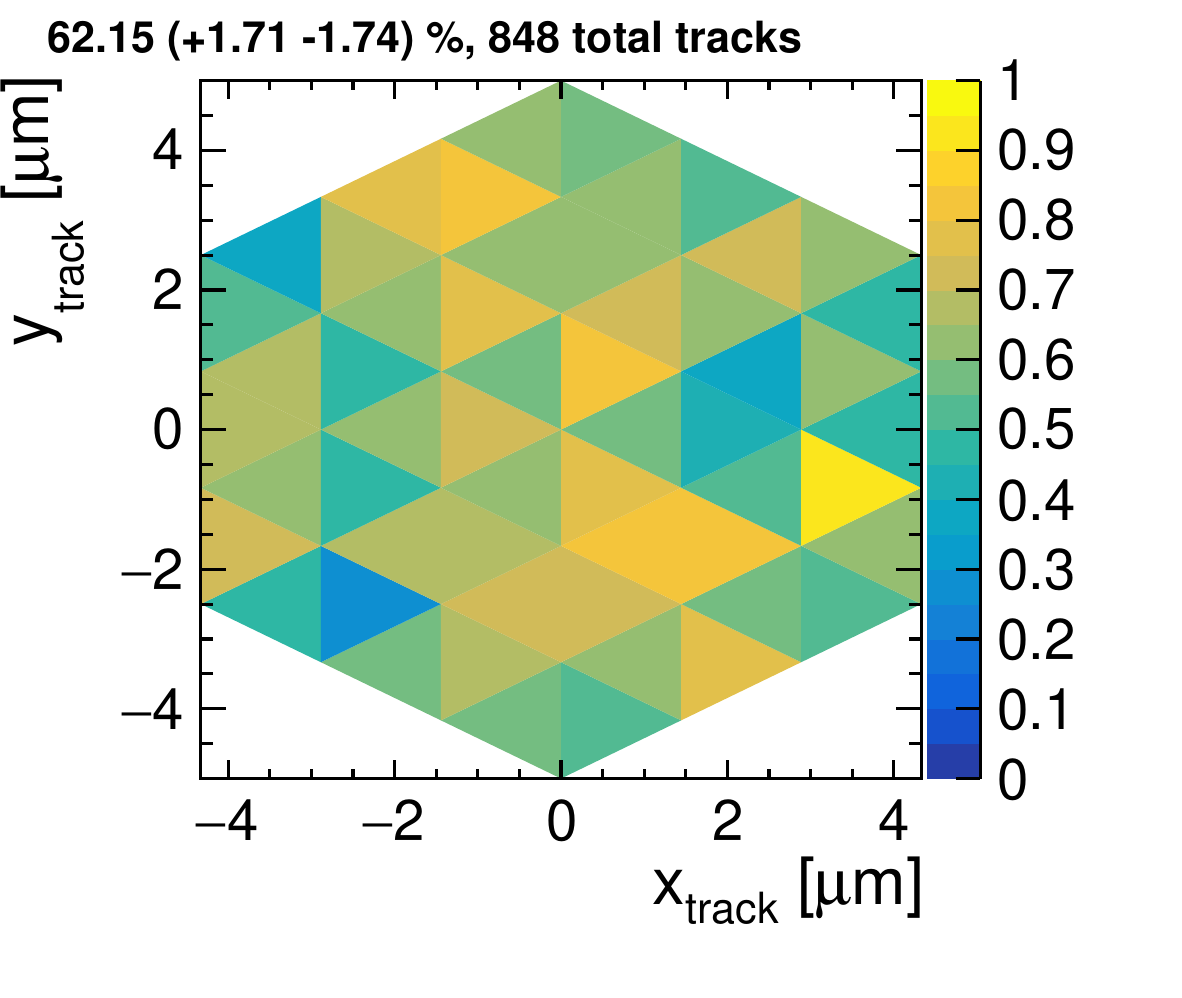}
        \caption{W03, matrix 3, \SI{8.66}{\um} pixel pitch, threshold \SI{66}{\electron}.}
    \end{subcaptionblock}
    \hfill
    \begin{subcaptionblock}[t]{0.495\textwidth}
        \centering
        \includegraphics[width=\textwidth]{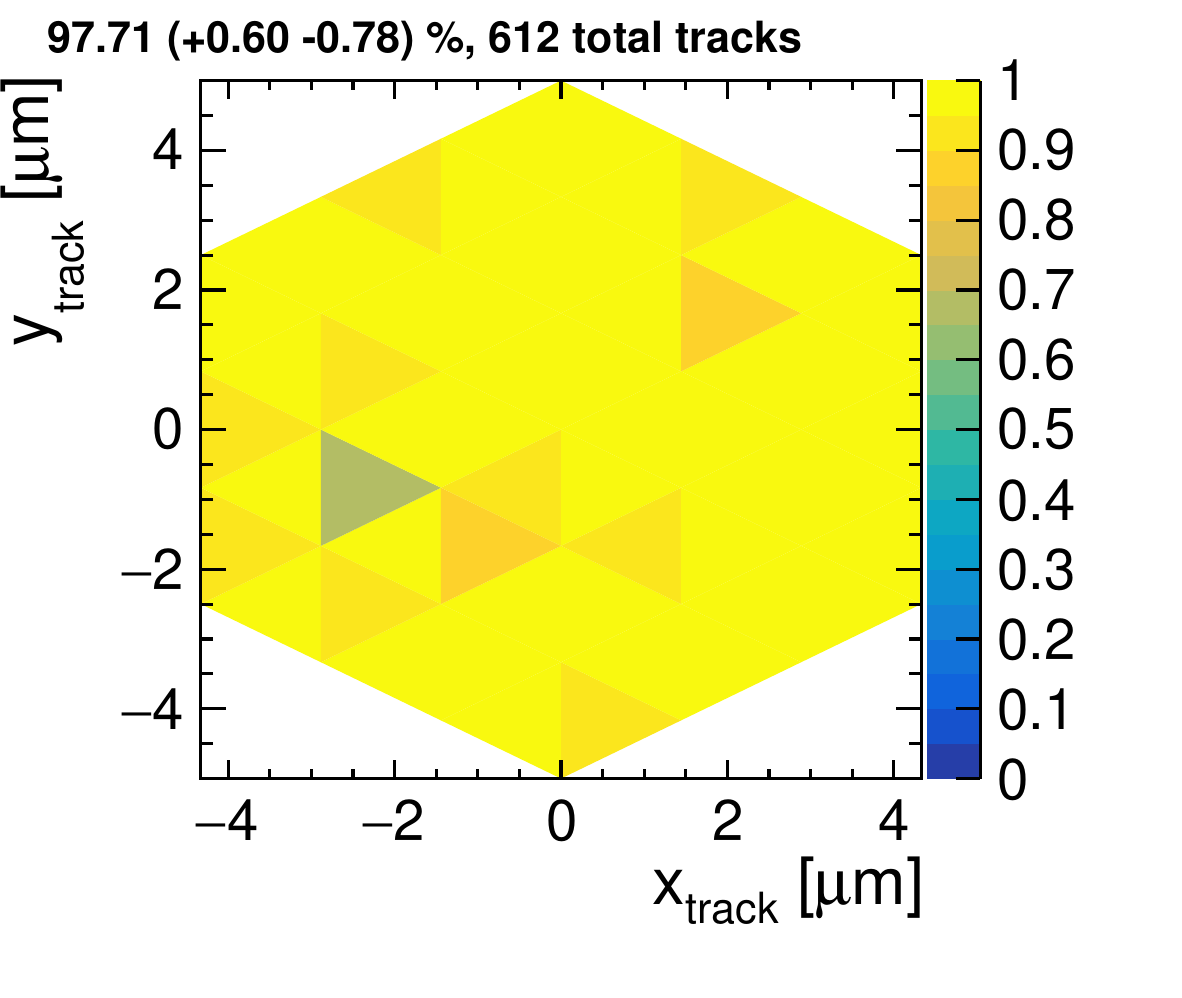}
        \caption{W15, matrix 3, \SI{8.66}{\um} pixel pitch, threshold \SI{51}{\electron}.}
    \end{subcaptionblock}
    \caption{In-pixel efficiency maps from a standard process (left) and a modified process (right) sample. Hits from all pixels are projected into one hexagonal pixel area. The color scale shows the efficiency for each triangular track position bin subdividing the hexagonal area on the x-/y-plane.}
    \label{fig:inpix_efficiency_8}
\end{figure}
\begin{figure}[h!]
    \centering
    \includegraphics[width=.5\linewidth]{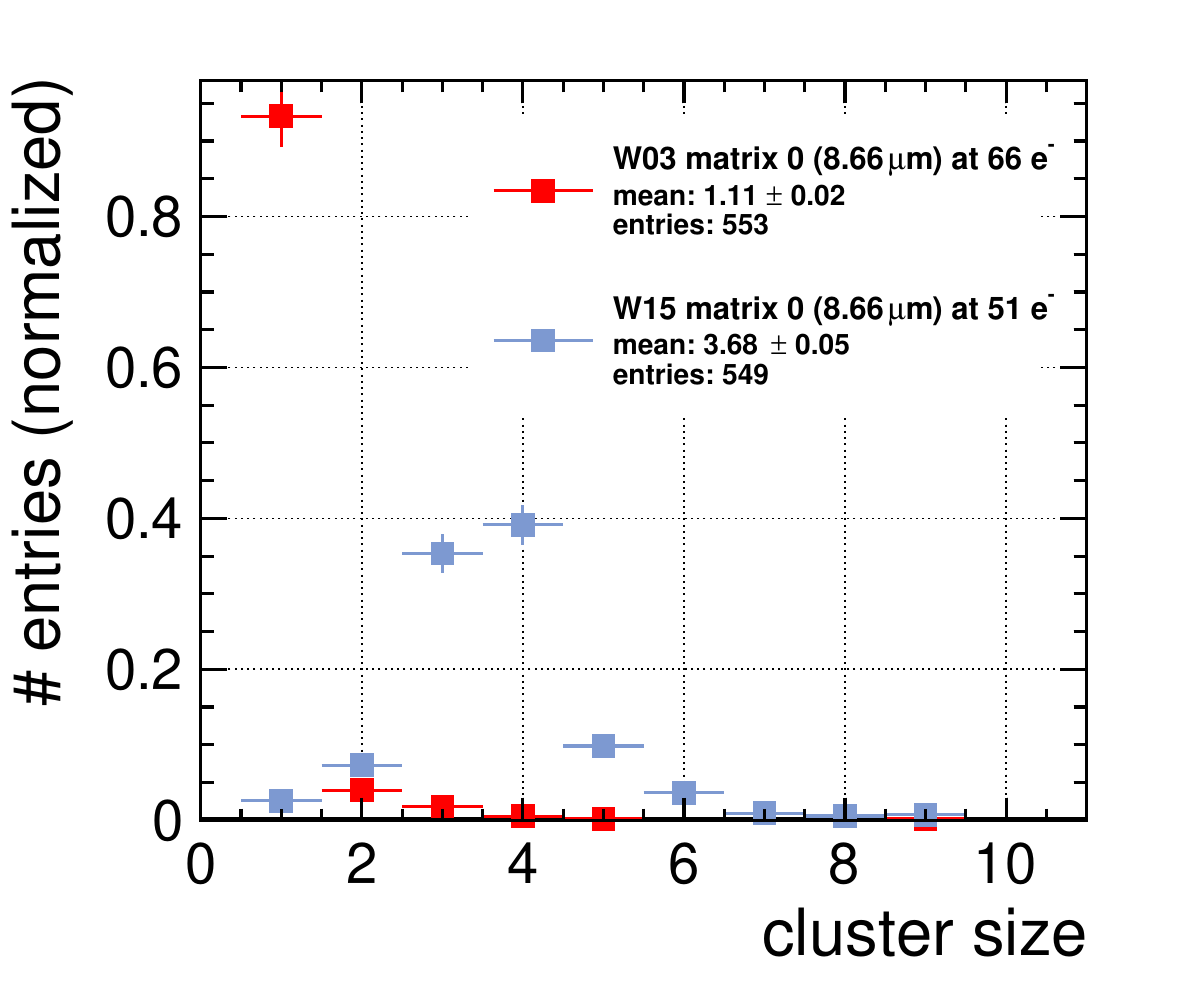}
    \caption{Cluster-size distributions of a \SI{8.66}{\um} pixel pitch matrix on W03 and W15.} 
    \label{fig:clstrsz_dist_8}
\end{figure}

\cref{fig:clstrsz_dist_20} shows normalized cluster-size distributions for \SI{20}{\um} pixel pitch matrix number \num{3} from the standard process sample W03 and the modified process sample W15.\\
On all wafers matrix \num{3} features pixels with a \SI{0.86}{\um} collection electrode, \SI{4.8}{\um} p-well opening and \SI{16.03}{\um} deep p-well opening.
\begin{figure}[b!]
    \centering
    \begin{subcaptionblock}[t]{0.495\textwidth}
        \centering
        \includegraphics[width=\textwidth]{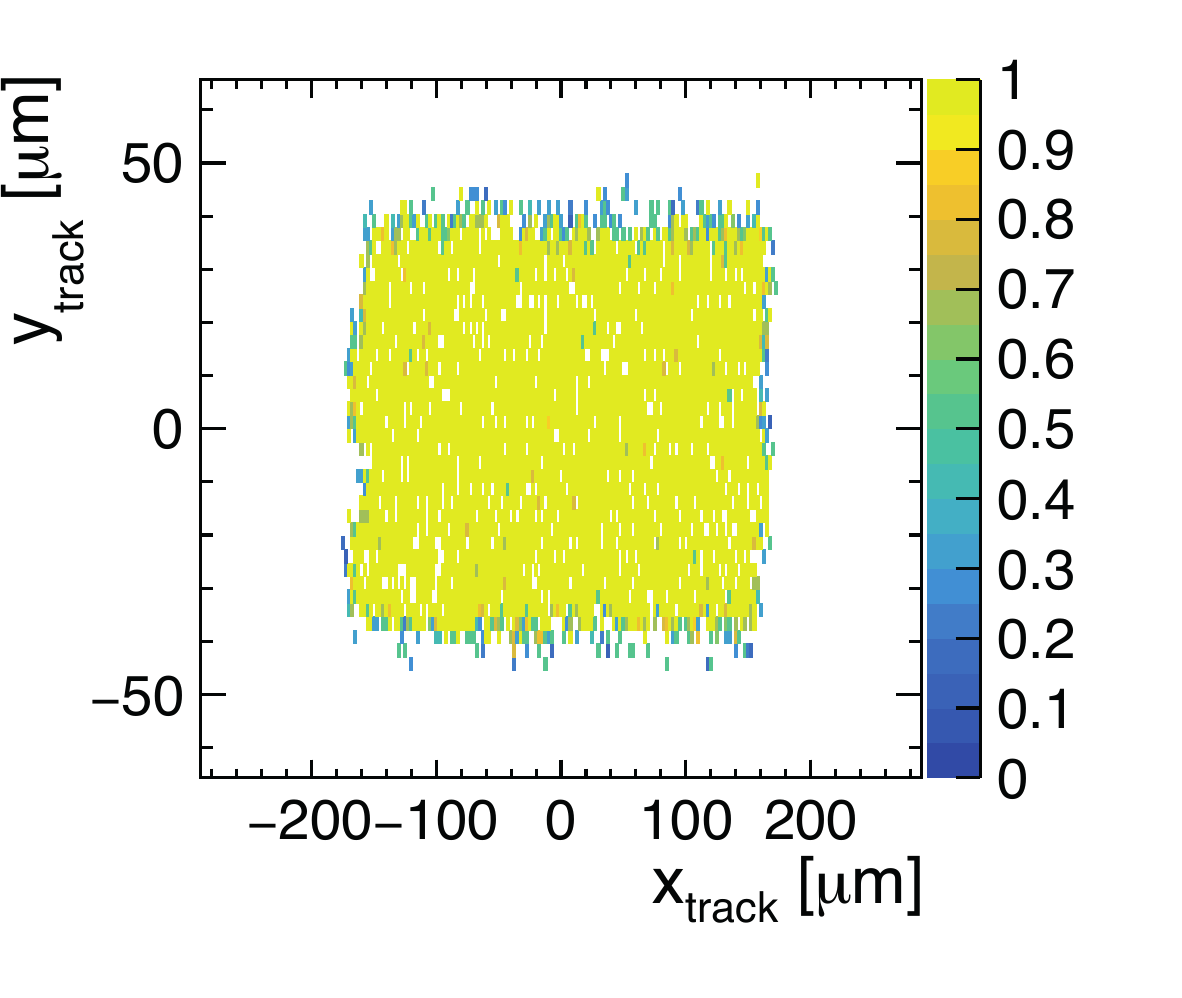}
        \caption{W03, matrix 3, \SI{20}{\um} pixel pitch.}
    \end{subcaptionblock}
    \hfill
    \begin{subcaptionblock}[t]{0.495\textwidth}
        \centering
        \includegraphics[width=\textwidth]{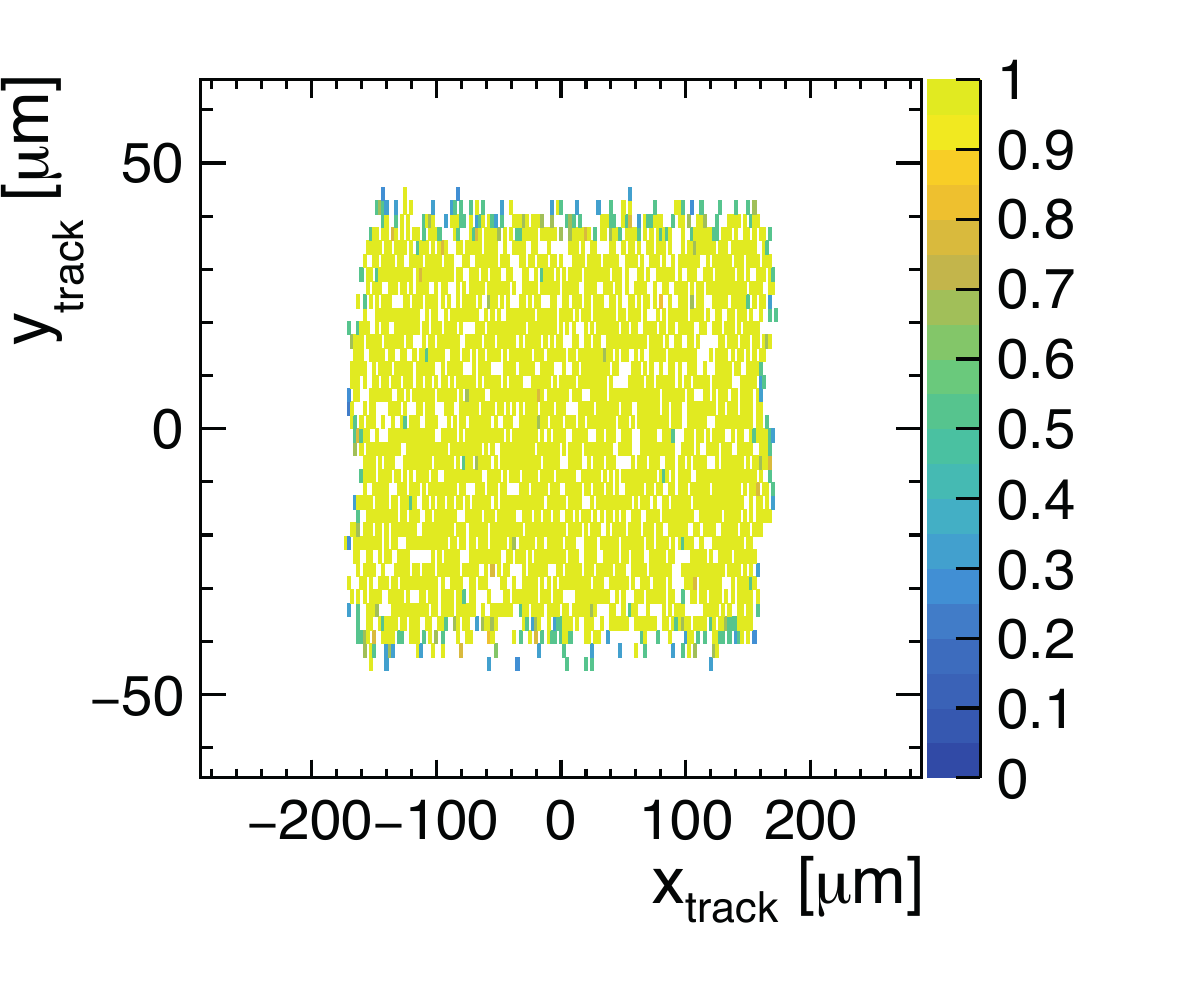}
        \caption{W15, matrix 3, \SI{20}{\um} pixel pitch.}
    \end{subcaptionblock}
    \caption{Matrix efficiency maps from a standard process (left) and a modified process (right) sample at a threshold of \SI{38}{\electron}. The color scale shows the efficiency for each track position bin on the x-/y-plane.}
    \label{fig:matrix_efficiency_20}
\end{figure}
The pixel edge of the same matrix on W15 is lined by an uninterrupted deep n-implant and has an \SI{18.8}{\um} extra-deep p-well opening, two modifications that accelerate charge collection by drift, reduce charge sharing and decrease cluster size.
Both cluster-size distributions follow similar trends and mean cluster size measures \num{1.83 \pm 0.02} pixels per cluster for W03 and \num{1.78 \pm 0.02} pixels per cluster for W15, with statistical uncertainty.\\
\cref{fig:matrix_efficiency_20} shows efficiency maps of matrix \num{3} on W03 and W15. 
The plots present a consistently efficient area of the matrix without visible structures introduced by characteristics of the chip or systematic features. 
A drop in efficiency along the edge is caused by charge sharing with inactive dummy-pixels surrounding the matrix edge making the sensitive area slightly larger than the 4 x 16 active pixels.\\
\cref{fig:inpix_efficiency_20} shows in-pixel efficiency maps of the same matrices with no visible characteristics of the pixel implant geometry or systematic features across the pixel area. 
For both samples matrix \num{3} is able to achieve fully efficient operation at a similar threshold of \SI{38}{\electron} without significant difference in cluster size or efficiency.\\
However, this observation changes for the smallest pixel pitch.
\cref{fig:matrix_efficiency_8} and \cref{fig:inpix_efficiency_8} show matrix and in-pixel efficiency maps of \SI{8.66}{\um} matrix \num{0} from the standard process sample W03 and the modified process sample W15, respectively.
Across all wafers matrix \num{0} features pixels with a \SI{0.86}{\um} collection electrode, \SI{3.0}{\um} p-well opening and \SI{4.8}{\um} deep p-well opening.
On W15 pixels in the same matrix have a continuous deep n-implant and a \SI{7.45}{\um} extra-deep p-well opening, two modifications that accelerate charge collection by drift, reduce charge sharing and decrease cluster size.\\
The efficiency plots present a consistently efficient area of the matrix as well as pixel area. 
With respect to the modified process sample W15, the standard process sample W03 experiences a decrease in efficiency of approximately \SI{35}{\percent} for \SI{8.66}{\um} pixel pitch matrix \num{0}.\\
Part of the cause of this decrease is a difference in physical threshold for
W03 at \SI{66}{\electron} compared to a threshold value of \SI{51}{\electron} for sample W15. 
A higher threshold reduces the observed mean cluster size, as lower-signal pixels in the border region of a cluster fall below threshold and remain undetected. 
If both samples operated fully efficient at the same threshold, the standard process sample W03 is expected to yield larger cluster size compared to the modified process sample W15 due to the larger amount of charge sharing (see \cref{sec:sensor_design}). 
With a \SI{15}{\electron} higher threshold for the standard process sample W03 the difference in cluster size between both process variants is reduced.\\
The cluster-size distribution in \cref{fig:clstrsz_dist_8} shows a 3 times higher mean cluster size for the modified process sample W15, compared to the the standard process sample W03.
A cause can be found from the implications of the very small \SI{8.66}{\um} pixel pitch.
A decrease in pitch towards smaller pixel size distributes a given amount of generated charge in the epitaxial layer to a larger number of pixels covering the area above the particle incidence. 
This results in an on average smaller signal on each pixel of a cluster, reducing both the efficiency and the measured cluster size.
The larger amount of charge sharing in the standard process enhances this reduction of efficiency and measured cluster size.
The geometry of the FASTPIX matrix imposes another effect that alters the registered cluster size in small pitch matrices. 
Another effect alters the registered cluster size in the \SI{8.66}{\um} pixel pitch matrices as some clusters span or even extend beyond the \num{4} pixel height of the matrix.
Clusters from tracks traversing a given matrix outside of the center line of the matrix experience edge effects. 
A fraction of their charge is deposited outside of the matrix, is not collected by active pixels and therefore not accounted for in a cluster size measurement. 
Consequently, a share of clusters get misidentified with a smaller size. 
To mitigate this effect the reconstructed local intercept position is used to only allow cluster seed pixel positions in the inner \num{2} x \num{14} pixels of a matrix. 
Comparing results for \SI{20}{\um} pixel pitch matrix \num{3} in \cref{fig:clstrsz_dist_20} and \SI{8.66}{\um} pixel pitch matrix \num{0} in \cref{fig:clstrsz_dist_8}, a shift in mean cluster size towards smaller values has been measured for standard process sample W03 matrix \num{0}, in contrast to the expected observation considering the differences in process variants.\\
The results show how the process modifications of the modified process sample W15 help to contain the charge within a single pixel and with that give more margin for efficient detector operation of small-pitch matrices.

\subsection{Spatial residuals}\label{sec:spatial_residuals}
\begin{figure}[h!]
    \centering
    \begin{subcaptionblock}[t]{0.495\textwidth}
        \centering
        \includegraphics[width=\textwidth]{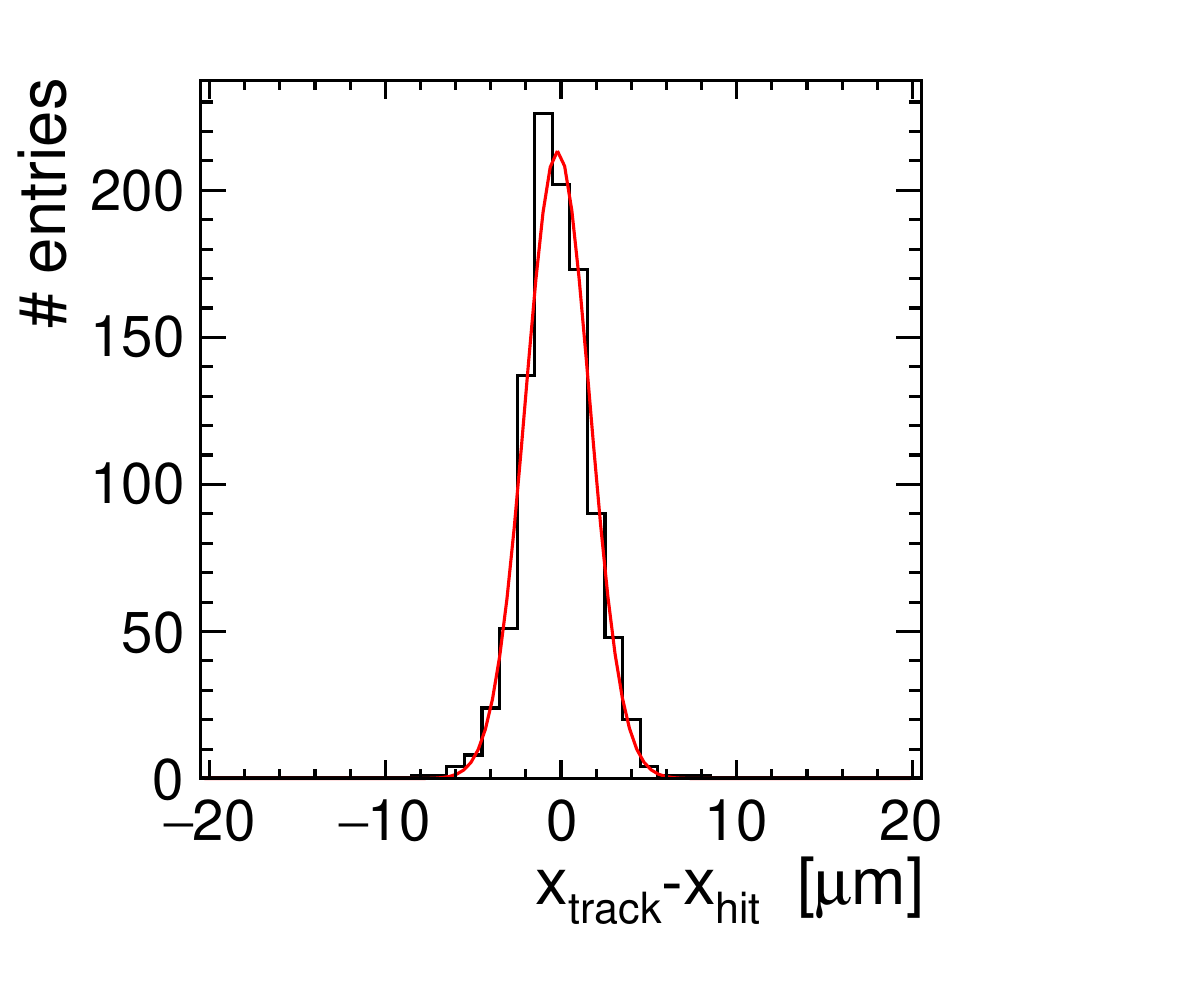}
        \caption{\SI{10}{\um} pitch matrix \num{1} at a threshold of \SI{62}{\electron}. The full distribution yields $\text{RMS} = \SI{1.89 \pm 0.04}{\um}$ and $\sigma_\text{fit, gaus} = \SI{1.82 \pm 0.05}{\um}$.}
        \label{fig:best_spatial_res}
    \end{subcaptionblock}
    \hfill
    \begin{subcaptionblock}[t]{0.495\textwidth}
        \centering
        \includegraphics[width=\textwidth]{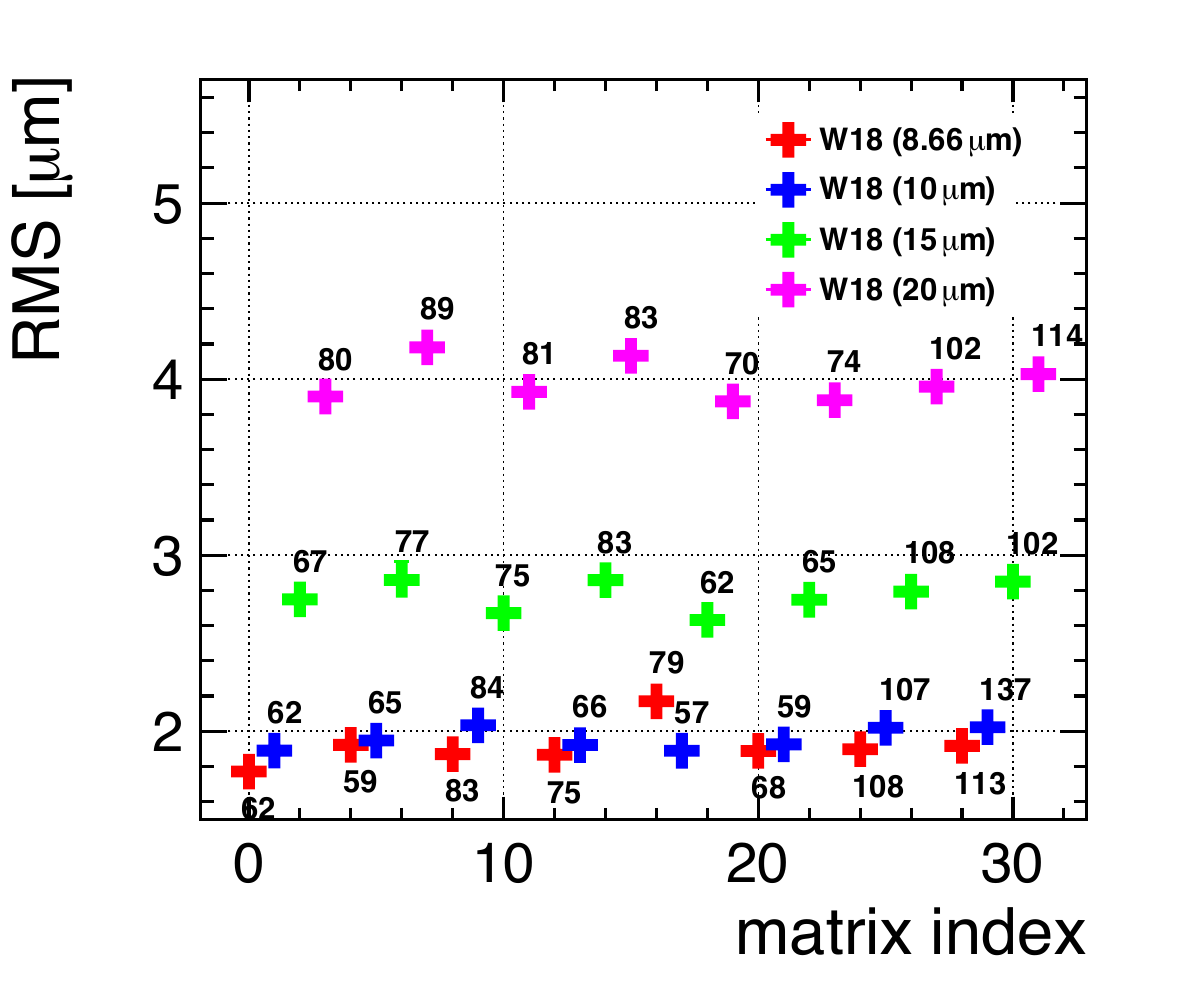}
        \caption{Overview of the width of spatial residuals along the x-axis measured with the W18 sample. The superscript of each data point gives the threshold in electrons for the respective matrix.}
        \label{fig:all_spatial_res}
    \end{subcaptionblock}
    \caption{Spatial residuals along the x-axis measured with the W18 sample.}
    \label{fig:spatial_res}
\end{figure}

An observable closely linked to cluster size is the width of the spatial residual distribution, shown for a \SI{10}{\um} pitch matrix from modified process sample W18 in \cref{fig:best_spatial_res}.\\
Similar to modified process sample W15, the sample W18 was produced in the modified process but with a higher-dose deep n-implant, which accentuates the achievable depletion and electric field configuration discussed in \cref{sec:sensor_design} even further.\\
An overview of the width of spatial residuals along the x-axis including all FASTPIX matrices of wafer \num{18} is shown in \cref{fig:all_spatial_res} in form of the RMS values between \SI{1.8}{\um} and \SI{4.2}{\um} against matrix number.
\\
Smaller pitch matrices achieve lower RMS values due to the higher granularity and since larger mean cluster size and a tendency for lower electron thresholds allow for a more precise reconstruction of the DUT hit position with respect to the interpolated track intersect.
Sub-threshold and matrix edge effects previously discussed in \cref{sec:clstrsz_eff} impact the amount of detected charge and with that have a negative influence on residual width for the smaller pixel pitches.\\
The \SI{8.66}{\um} pixel pitch matrices achieve a spatial resolution down to \SI{1}{\um} after unfolding the telescope tracking resolution of \SI{1.7}{\um} \cite{AlipourTehrani:2270788}.

\subsection{Time residuals}\label{sec:time_residuals}
\begin{figure}[t]
    \centering
    \begin{subcaptionblock}[t]{0.495\textwidth}
        \centering
        \includegraphics[width=\textwidth]{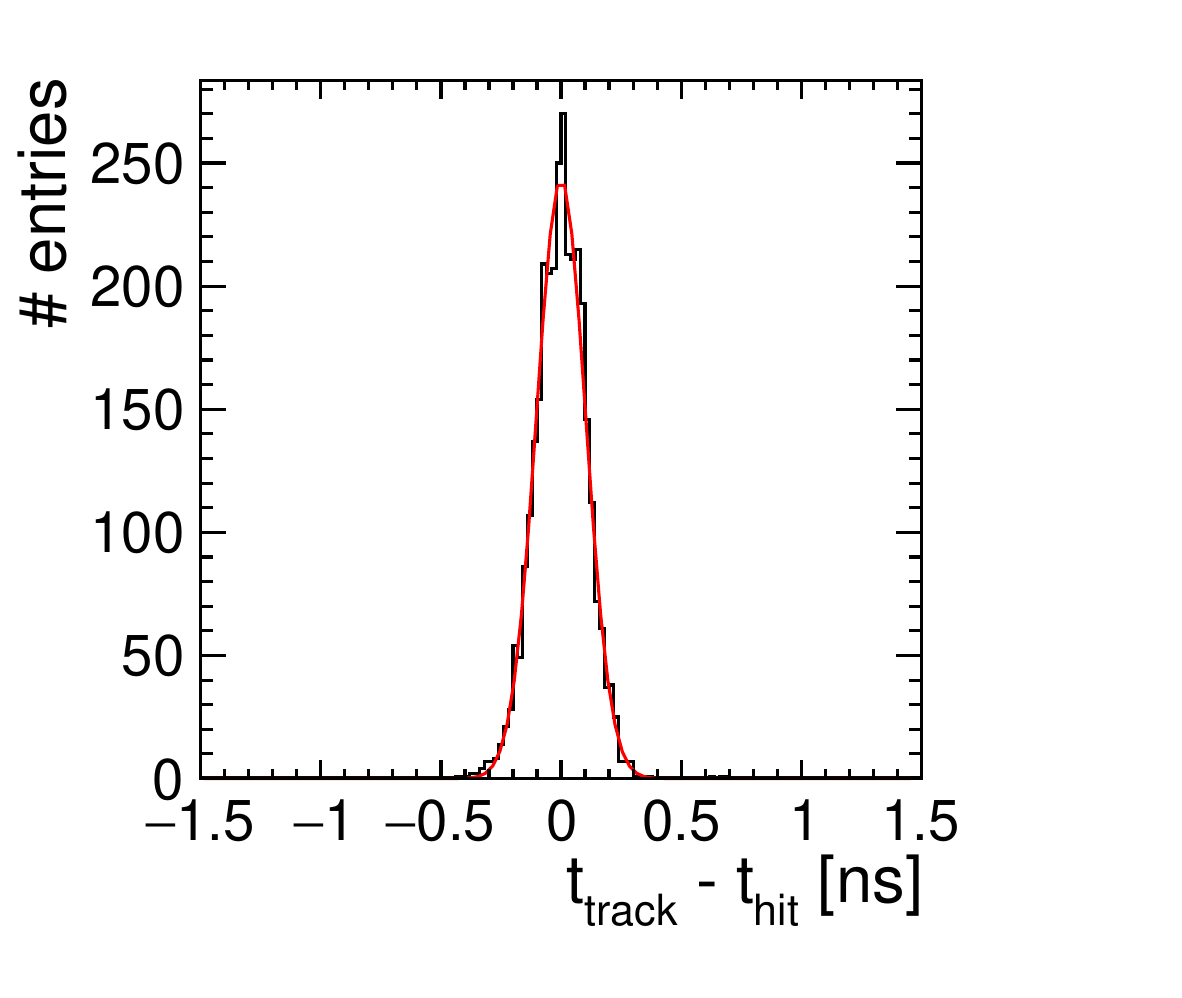}
        \caption{Time residual from matrix \num{23} at a threshold of \SI{74}{\electron}. The full distribution yields $\text{RMS} = \SI{107 \pm 2}{\ps}$, $\text{RMS}_{\SI{99.7}{\percent}} = \SI{103 \pm 0.3}{\ps}$ and $\sigma_\text{fit, gaus} = \SI{102 \pm 1}{\ps}$. The errors represent statistical uncertainties.}
        \label{fig:best_time_res}
    \end{subcaptionblock}
    \hfill
    \begin{subcaptionblock}[t]{0.495\textwidth}
        \centering
        \includegraphics[width=\textwidth]{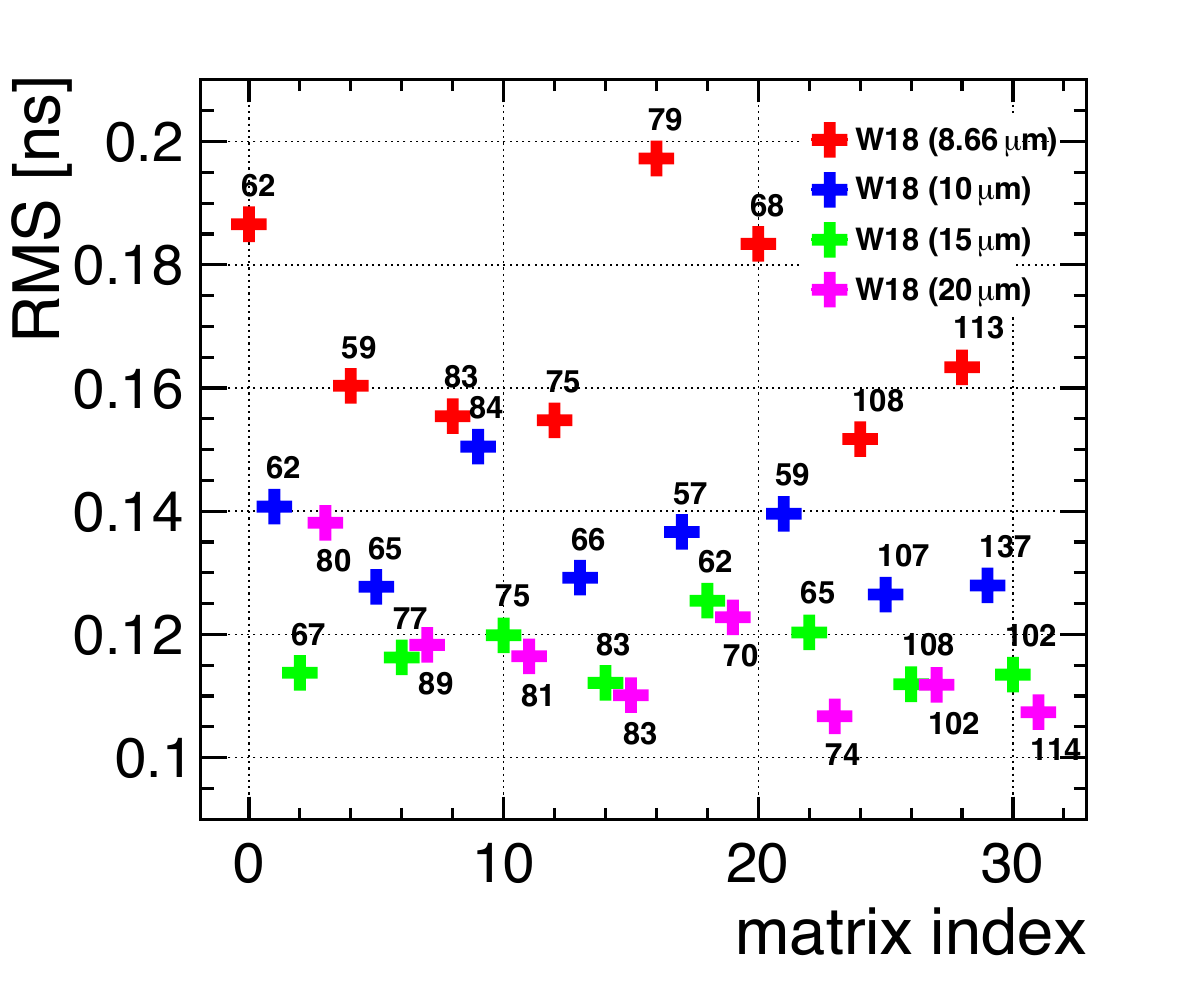}
        \caption{Overview of the width of time residual distributions from all W18 matrices. The superscript of each data point gives the threshold in electrons for the respective matrix.}
        \label{fig:all_time_res}
    \end{subcaptionblock}
    \caption{Time residuals measured with the W18 sample.}
    \label{fig:time_res}
\end{figure}
\begin{figure}[h!]
    \centering
    \begin{subcaptionblock}[t]{0.495\textwidth}
        \centering
        \includegraphics[width=\textwidth]{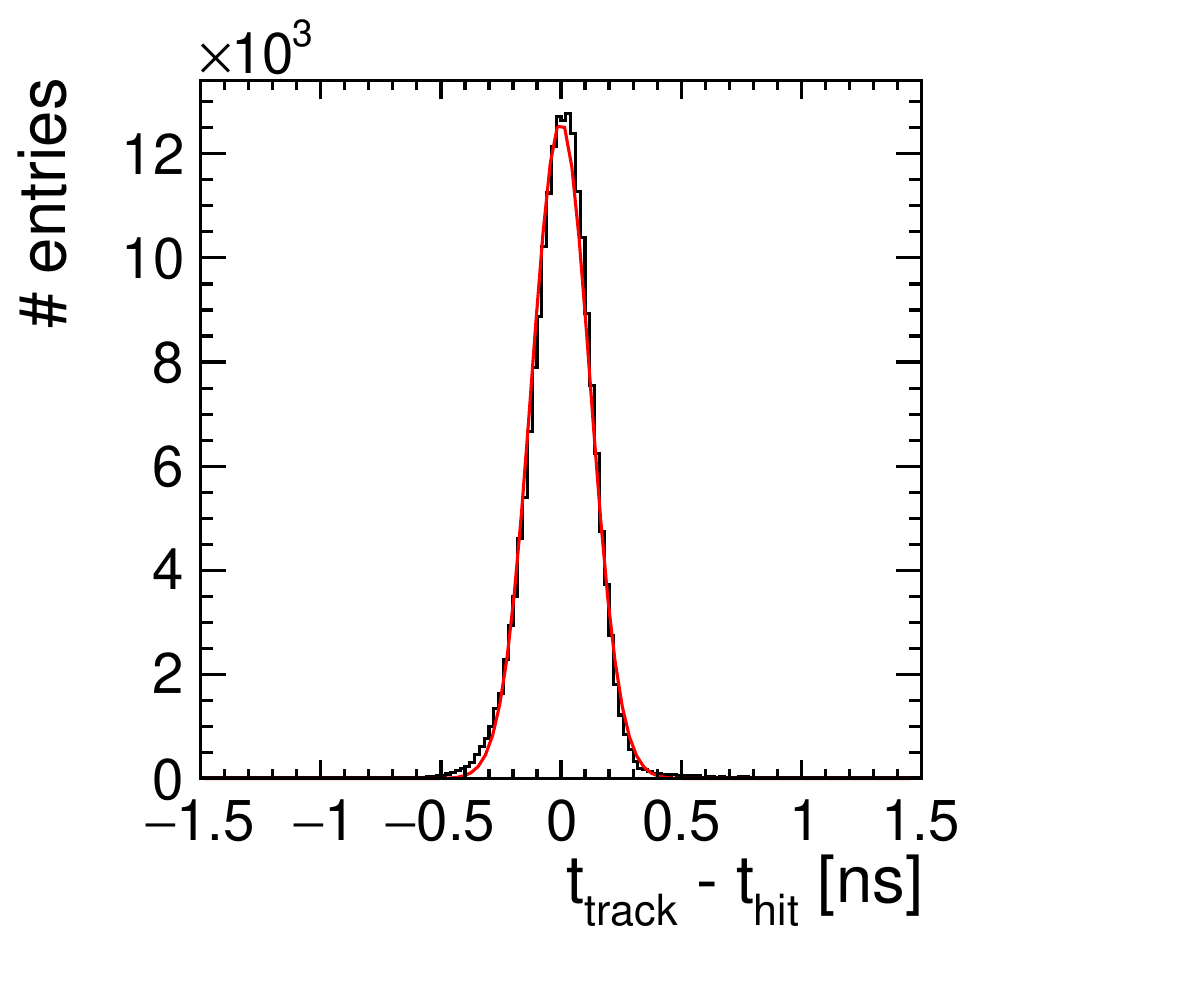}
        \caption{Matrix \num{1}, \SI{10}{\um} pixel pitch at a threshold of \SI{62}{\electron}. The distribution yields $\text{RMS} = \SI{140.7 \pm 0.2}{\ps}$, $\text{RMS}_{\SI{99.7}{\percent}} = \SI{128 \pm 0.2}{\ps}$ and $\sigma_\text{fit, gaus} = \SI{121.5 \pm 0.2}{\um}$.}
    \end{subcaptionblock}
    \hfill
    \begin{subcaptionblock}[t]{0.495\textwidth}
        \centering
        \includegraphics[width=\textwidth]{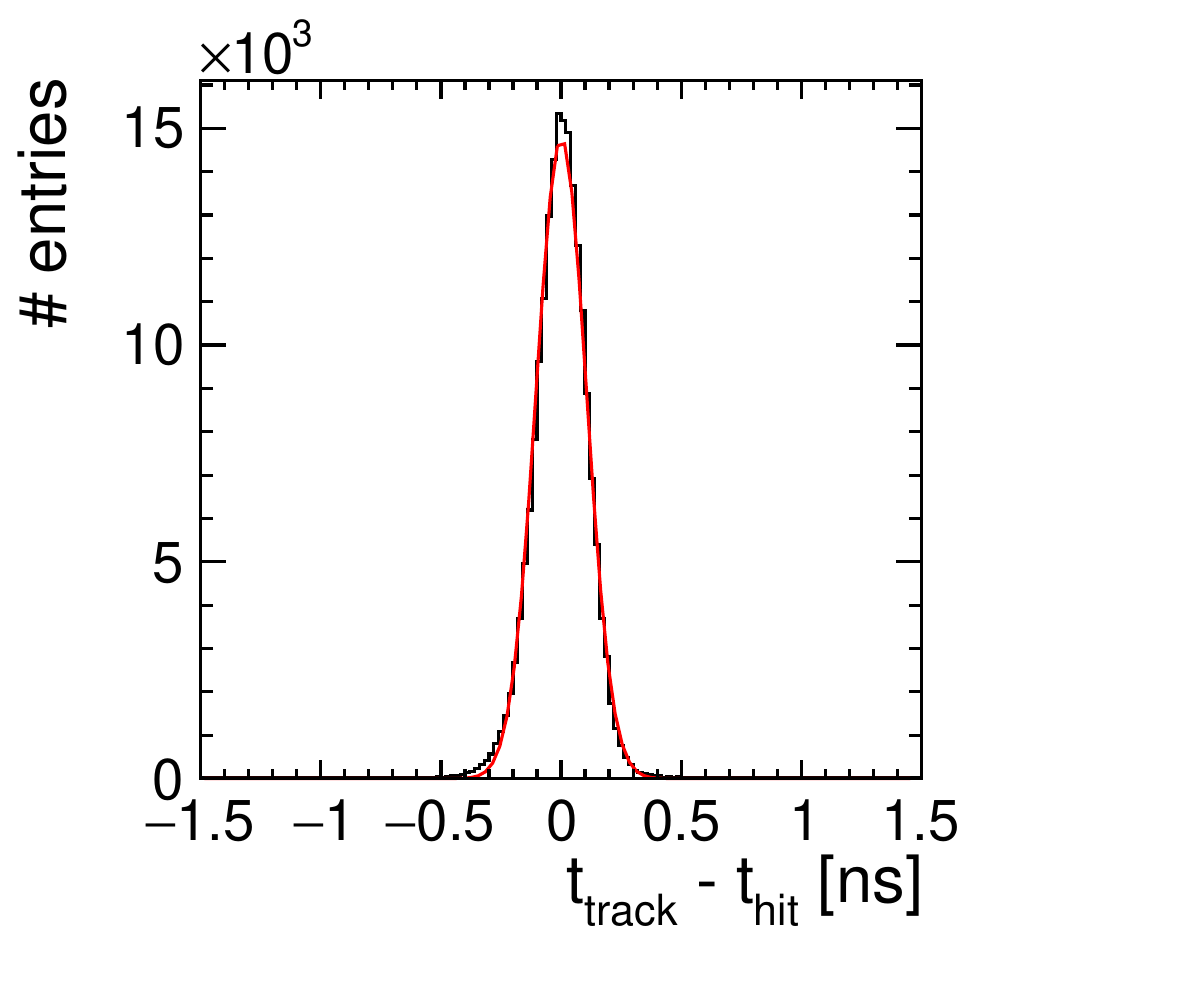}
        \caption{Matrix \num{3}, \SI{20}{\um} pixel pitch at a threshold of \SI{80}{\electron}. The distribution yields $\text{RMS} = \SI{138.2 \pm 0.2}{\ps}$, $\text{RMS}_{\SI{99.7}{\percent}} = \SI{109 \pm 0.2}{\ps}$ and $\sigma_\text{fit, gaus} = \SI{104.4 \pm 0.2}{\um}$.}
    \end{subcaptionblock}
    \caption{Time residual distributions for two matrices of modified process sample W18. The errors represent statistical uncertainties.}
    \label{fig:time_res_by_pitch}
\end{figure}
\begin{figure}[h!]
    \centering
    \begin{subcaptionblock}[t]{0.495\textwidth}
        \centering
        \includegraphics[width=\textwidth]{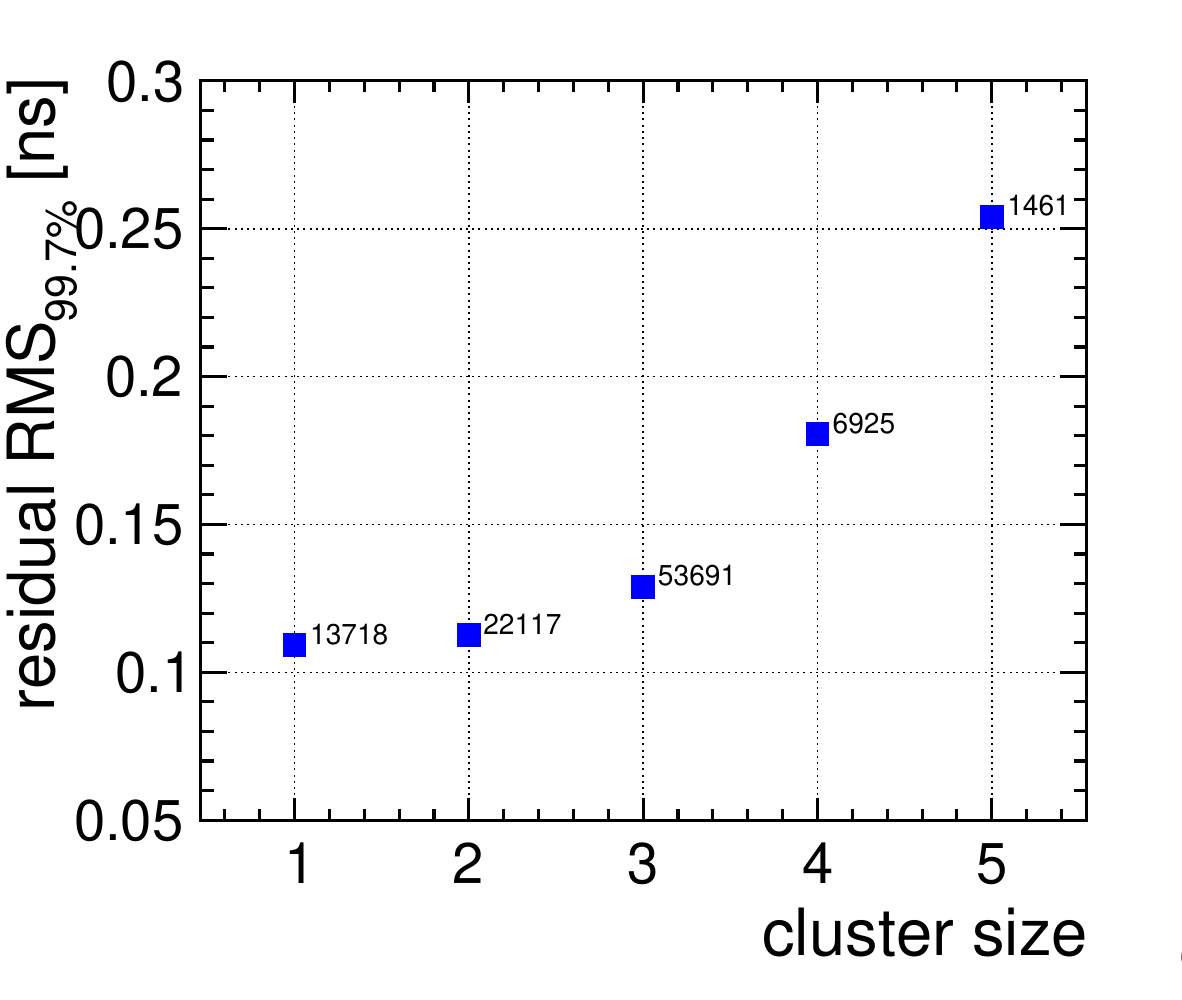}
        \caption{Matrix \num{1}, \SI{10}{\um} pixel pitch at a threshold of \SI{62}{\electron}. The distribution for single-pixel clusters yields: \SI{109.2 \pm 0.1}{\ps} RMS$_{\SI{99.7}{\percent}}$}
    \end{subcaptionblock}
    \hfill
    \begin{subcaptionblock}[t]{0.495\textwidth}
        \centering
        \includegraphics[width=\textwidth]{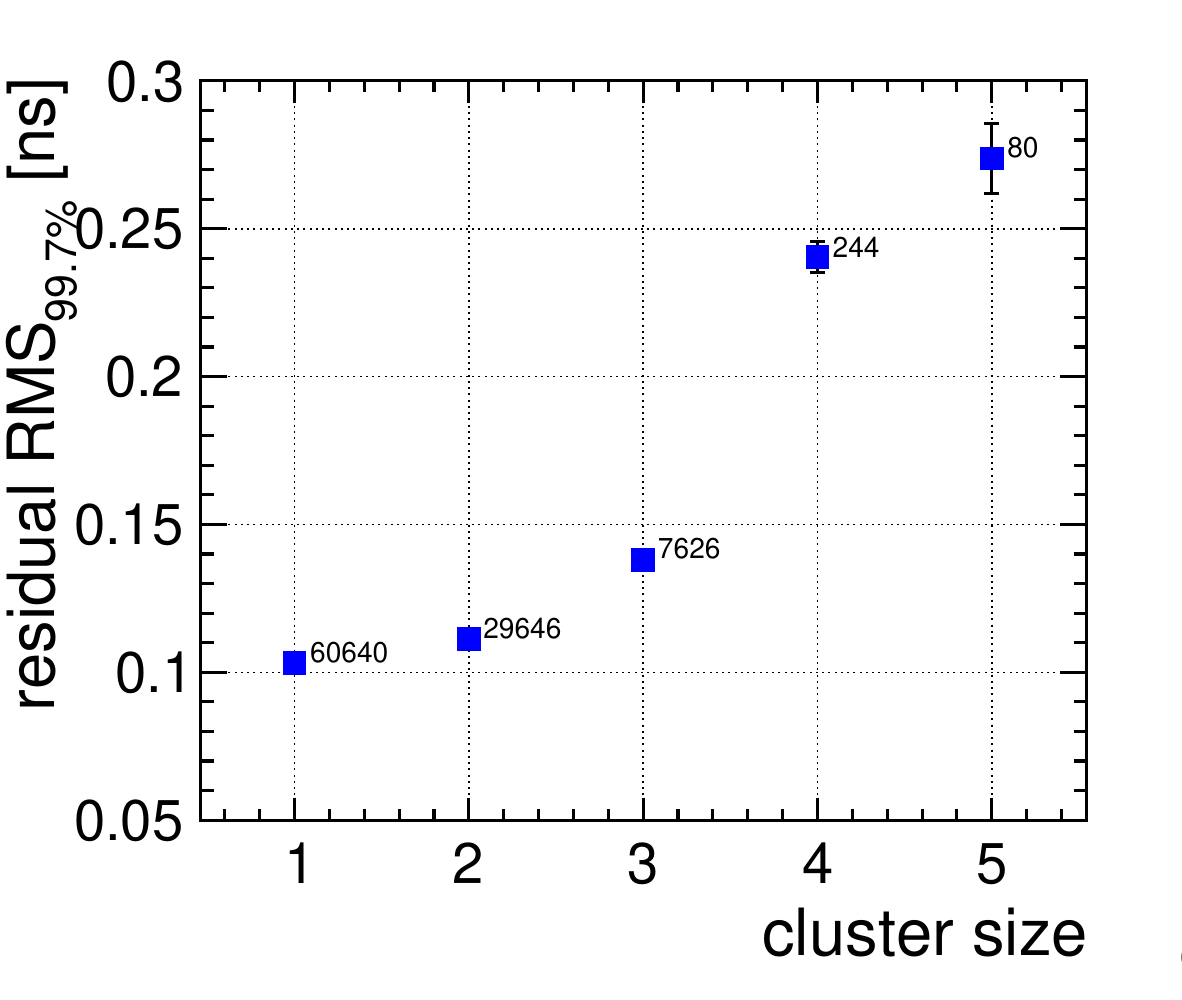}
        \caption{Matrix \num{3}, \SI{20}{\um} pixel pitch at a threshold of \SI{80}{\electron}. The distribution for single-pixel clusters yields: \SI{103.2 \pm 0.1}{\ps} RMS$_{\SI{99.7}{\percent}}$}
    \end{subcaptionblock}
    \caption{Time residual width (RMS$_{\SI{99.7}{\percent}}$) per cluster size for two matrices of modified process sample W18. The error bars represent the statistical uncertainties. The superscript of each data point gives the number of entries in the subset of the underlying time residual distribution.}
    \label{fig:time_res_by_clstrz}
\end{figure}

Timing performance is evaluated for the modified process with higher-dose deep n-implant. 
The higher dose n-implant of the W18 sample accentuates the gradient in the doping profile and the lateral component of the field, driving charges from the pixel border towards the collection electrode maximizing the seed pixel signal and reducing time-walk effects.
The width of the time residual is shown in \cref{fig:best_time_res} for \SI{20}{\um} pixel pitch matrix \num{23}. 
The RMS yields \SI{107 \pm 2}{\ps} and is slightly higher than the width of a Gauss fit $\sigma = \SI{102 \pm 1}{\ps}$ performed on the same set of data. 
The fit is in good agreement with the RMS$_{\SI{99.7}{\percent}} = $ \SI{103 \pm 0.3}{\ps}, for which the outmost \SI{.3}{\percent} of outliers are excluded from the calculation.\\
For a selection of FASTPIX matrices of modified process sample W18 a high-statistic set of data is available, exhibiting less contribution of non-Gaussian tails in the time residual. 
With \num{\approx 30} times more entries than in e.g. \cref{fig:best_time_res} the impact of pixel pitch is illustrated by \cref{fig:time_res_by_pitch}.
Both matrices differ in pixel pitch while the other pixel design modification parameters are equal, except deep p-well and extra-deep p-well opening which both have a specific size for each pixel pitch in the quadrants of the FASTPIX chip (see \cref{fig:parameter_space}).
The \SI{20}{\um} pixel pitch matrix achieves \SI{15}{\percent} better timing performance compared to the \SI{10}{\um} pixel pitch.\\
\\
The timewalk-correction approach presented in \cref{sec:timewalk_corr} enables an investigation of the contribution of seed pixel signals of different size clusters to the time measurement precision of the chip. 
In case of a cluster size \num{> 1} less signal is collected by the seed pixel, giving way to time-walk effects.
Moreover, for large cluster size it is more likely that a particle hit falls on the corner of the pixel, where charge collection is slower and exhibits more spread in collection speed.\\
\cref{fig:time_res_by_clstrz} shows the width of time residuals from subsets of data sorted by cluster size. Time difference values $\Delta \text{t} = \text{t}_\text{DUT} - \text{t}_\text{MCP}$ for cluster sizes \num{<= 5} are sorted in exclusive data sets while clusters \num{> 5} are grouped in a combined set.
In both cases, independent of pitch, single pixel clusters reach the best performance with a higher probability of collecting the majority of deposited charge in a singular pixel. 
Doubling the pixel pitch yields an improvement of \SI{6}{\percent}, from an RMS$_{\SI{99.7}{\percent}}$ of \SI{109}{\ps} for the \SI{10}{\um} pixel pitch matrix to \SI{103}{\ps} in case of the \SI{20}{\um} pixel pitch matrix. 
Large pixel pitch is in favor of timing performance as a large part of the signal charge is collected by the seed pixel, reducing time-walk effects.\\
\\
\cref{fig:all_time_res} shows an overview of the time residual RMS of all FASTPIX matrices based on a low-statistic set of data.
The impact of different pixel design modifications can be discussed by isolating all matrices of equal pitch from the overview in \cref{fig:all_time_res}.\\ 
The data points for matrices with \SI{20}{\um} pixel pitch are colored in pink.
In this subset the baseline is matrix \num{3} which achieves a time residual RMS of \SI{138}{\ps}. 
Following \cref{fig:parameter_space}, this matrix features pixels with a \SI{0.86}{\um} collection electrode, \SI{4.8}{\um} p-well opening, \SI{16.03}{\um} deep p-well opening and \SI{18.8}{\um} extra-deep p-well opening. 
The pixel edge is lined by an uninterrupted deep n-implant without modifications of the deep n-implant geometry.\\
For matrix \num{7} a hexagonal \SI{2.6}{\um} corner gap gets added to the deep n-implant while the other parameters remain unchanged. 
The gap increases the strength of the lateral electric field in the corner regions of the pixel which accelerates charge collection by drift, reduces charge sharing and maximizes the seed pixel signal. 
The improved timing performance yields a time residual RMS of \SI{118}{\ps}.\\
Another variation of the deep n-implant is found in matrix \num{23} where the corner gap is changed to a \SI{3.8}{\um} triangular geometry. 
The larger gap produces an even more pronounced pull away from the pixel corner towards the collection electrode and further narrows the spread in timing performance, resulting in a time residual RMS of \SI{106}{\ps}.\\
Matrix \num{19} differs from matrix \num{3} in terms of p-well opening size, while the remaining parameters stay the same. 
Reducing the area of the p-well benefits a uniform depletion over the full pixel area, favors charge collection but also cuts into the area available for in-pixel circuitry and its complexity.
The smaller p-well area of matrix \num{19} with a \SI{6.0}{\um} p-well opening achieves a time residual RMS of \SI{123}{\ps}.\\
The collection-electrode size is another parameter that is changed throughout these eight matrices. 
Increasing the size of the collection electrode has similar beneficial effects as the change in p-well opening but comes with a penalty for signal-to-noise ratio and power consumption.
Comparing matrix \num{19} (\SI{123}{\ps}) and \num{27} (\SI{112}{\ps}) shows the impact of a change in collection electrode size from \SI{0.86}{\um} to \SI{2.0}{\um} with an uninterrupted deep n-implant geometry. 
Comparing matrix \num{7} (\SI{118}{\ps}) and \num{15} (\SI{110}{\ps}) yield slightly less gain in timing performance from the same difference in collection electrode size with a hexagonal \SI{2.6}{\um} corner gap.\\
In comparison with matrix \num{3}, an approximate doubling of the collection electrode size and a \SI{\approx 50}{\percent} increase in p-well opening size brought improvements in timing performance of \SI{\approx 9}{\percent} and \SI{\approx 12}{\percent}, respectively.
The larges improvement was observed with the addition of corner gaps, especially the triangular corner gaps could yield a decrease in time residual width of \SI{\approx 17}{\percent} compared to the baseline matrix \num{3}.

\section{Conclusions}\label{sec:conclusion}
FASTPIX was successfully integrated in the Timepix3 telescope, the Caribou DAQ and Corryvreckan-based reconstruction and analysis framework. 
Large data sets allowed the comparison of the various process modifications, sensor layouts and pixel pitches.\\
The process modifications have shown to be essential to maintain efficient operation for small pixel pitch matrices. 
FASTPIX reaches a spatial resolution down to \SI{1}{\um} and $\mathcal{O}(\SI{100}{\ps})$ timing precision for the modified process with higher-dose deep n-implant.\\
The impact of sensor design optimization was presented in context of timing performance results for \SI{20}{\um} pixel pitch matrices from a modified process sample with higher-dose deep n-implant.
The largest relative improvement was observed for additional triangular corner gaps, followed by improvements from a change in collection electrode and p-well opening size.\\
The obtained FASTPIX results with \SI{180}{\nm} technology demonstrate the large potential of targeted sensor process and TCAD simulation-based design optimisations for the detector performance. 
Similar optimizations are also applicable in \SI{65}{\nm} process technology, which enables higher circuit density and thereby allows to include advanced readout circuitry for precise timing inside a larger pixel matrix.

\section*{Acknowledgments}
\begin{description}
\item[Authorship contribution statement:] J.B.: Investigation, Software, Visualization and Writing-Original Draft; E.B.: Investigation, Software and Visualization; D.D.: Investigation, Supervision, Writing-Review and Editing; K.D.: Investigation; T.K.: Investigation; M.M.: Investigation; W.S.: Conceptualization and Funding Acquisition; P.S.: Investigation; M.V.: Investigation. All authors have read and agreed to the published version of the manuscript.
\item[Declaration of competing interest:] The authors declare that they have no known competing financial interests or personal relationships that could have appeared to influence the work reported in this paper.
\item[Data availability statement:] The data supporting the findings of this study are available within the article.
\item[Funding:] This project has received funding from the ATTRACT project, funded by the European Commission under grant agreement 777222.\\This work has been sponsored by the Wolfgang Gentner Programme of the German Federal Ministry of Education and Research (grant no. 05E15CHA).
\item[Infrastructure:] We would like to gratefully acknowledge the CERN SPS beam physicists and accelerator staff for the reliable and efficient test beam operation. We thank Eraldo Oliveri for his help with providing and operating the MCP-PMT.
\end{description}
\printbibliography
\end{document}